\documentclass[11pt,tightenlines,eqsecnum,floats,aps,amsmath,amssymb,nofootinbib,prd,shownopacs,floatfix]{revtex4}
\usepackage{graphicx}
\usepackage{epstopdf}
\usepackage{latexsym}
\usepackage{amssymb}
\usepackage{amsmath}
\usepackage{color}
\usepackage{mathrsfs}
\usepackage{xparse}
\usepackage{float}
\usepackage{mathtools}
\usepackage[center]{subfigure}

\begin{document}
\renewcommand\arraystretch{2}
\newcommand{\bq}{\begin{equation}}
\newcommand{\eq}{\end{equation}}
\newcommand{\bqn}{\begin{eqnarray}}
\newcommand{\eqn}{\end{eqnarray}}
\newcommand{\nb}{\nonumber}
\newcommand{\lb}{\label}
\newcommand{\cb}{\color{blue}}
\newcommand{\cc}{\color{cyan}}
\newcommand{\cm}{\color{magenta}}
\newcommand{\rc}{\rho^{\scriptscriptstyle{\mathrm{I}}}_c}
\newcommand{\rd}{\rho^{\scriptscriptstyle{\mathrm{II}}}_c} 
\NewDocumentCommand{\evalat}{sO{\big}mm}{%
  \IfBooleanTF{#1}
   {\mleft. #3 \mright|_{#4}}
   {#3#2|_{#4}}%
}
\newcommand{\PRL}{Phys. Rev. Lett.}
\newcommand{\PL}{Phys. Lett.}
\newcommand{\PR}{Phys. Rev.}
\newcommand{\CQG}{Class. Quantum Grav.}
\newcommand{\parallelsum}{\mathbin{\!/\mkern-5mu/\!}}

\title{On a close relationship between the dressed metric and the hybrid approach to perturbations in effective loop quantum cosmology}

\author{Bao-Fei Li$^{1,2}$}
\email{baofeili1@lsu.edu}
\author{Parampreet Singh$^1$}
\email{psingh@lsu.edu}
\affiliation{
$^{1}$ Department of Physics and Astronomy, Louisiana State University, Baton Rouge, LA 70803, USA\\
$^{2}$ Institute for Theoretical Physics $\&$ Cosmology, Zhejiang University of Technology, Hangzhou, 310032, China\\}

\begin{abstract}
The dressed metric and the hybrid approach to perturbations are the two main approaches to capture the effects of quantum geometry in the primordial power spectrum in loop quantum cosmology. Both consider Fock quantized perturbations over a loop quantized background and result in very similar predictions except for the modes which exit the horizon in the effective spacetime in the Planck regime. Understanding precise relationship between both approaches has so far remained obscured due to differences in construction and  technical assumptions.  We explore this issue at the classical and effective spacetime level for linear perturbations, ignoring backreaction, which is the level at which practical computations of the power spectrum in both of the  approaches have so far been performed. We first show that at the classical level both the approaches  lead to the same Hamiltonian up to the second order in perturbations and result in the same classical mass functions in the Mukhanov-Sasaki equation on the physical solutions. At the effective spacetime level, the difference in phenomenological predictions between the two approaches in the Planck regime can be traced to whether one uses the Mukhanov-Sasaki variable $Q_{\vec k}$ (the dressed metric approach) or its rescaled version $\nu_{\vec k}=aQ_{\vec k}$ (the hybrid approach) to write the Hamiltonian of the perturbations, and associated polymerization ambiguities. It turns out that if in the dressed  metric approach one chooses to work with $\nu_{\vec{k}}$, the effective mass function can be written exactly as in the hybrid approach, thus leading to identical phenomenological predictions in all regimes.  
Our results explicitly show that the dressed metric and the hybrid approaches for linear perturbations, at a practical computational level,  can be seen as two sides of the same coin.
\end{abstract}

\maketitle

\section{Introduction}
\label{sec:introduction}
\renewcommand{\theequation}{1.\arabic{equation}}\setcounter{equation}{0}

Loop quantum cosmology (LQC) \cite{Ashtekar:2011ni}, an approach to explore quantization of cosmological spacetimes based on loop quantum gravity (LQG), has emerged in recent years as an elaborate framework to investigate the impact of non-perturbative quantum gravitational effects in the very early universe. Unlike previous attempts in quantum cosmology, LQC is based on a (discrete) quantum geometry  which replaces the classical differential geometry near the Planck scale. Thanks to this quantum geometry, the spacetime curvature is bounded in the Planck regime and the big bang singularity is replaced by a big bounce \cite{Ashtekar:2006rx,Ashtekar:2006uz,Ashtekar:2006wn, Ashtekar:2007em}. In the last two decades, significant progress has been made in LQC to understand the quantum evolution of the universe across cosmological singularities and 
the imprints of quantum geometry on linear perturbations around the background effective quantum spacetime.  In particular, 
the resolution of curvature singularities via a non-singular quantum bounce has been proved  to be a generic feature in both isotropic and anisotropic spacetimes \cite{Singh:2009mz,Singh:2011gp,Singh:2014fsy,Saini:2017ggt,Saini:2017ipg}. But so far there has been no complete treatment of various modes of cosmological perturbations 
encoding quantum geometric effects in a sense where perturbations are also loop quantized.\footnote{Two attempts in this direction are the deformed algebra approach \cite{Bojowald:2008gz,Cailleteau:2012fy,Cailleteau:2011kr} and the separate universe approach valid for long wavelength modes \cite{Wilson-Ewing:2015sfx}.}
In order to understand phenomenological implications of quantum geometry in the cosmic microwave background (CMB), a pragmatic strategy is to consider Fock quantized linear perturbations over a loop quantized background (see for reviews \cite{Agullo:2016tjh, Ashtekar:2015dja, ElizagaNavascues:2016vqw, Wilson-Ewing:2016yan}). There are two approaches in this direction -- the dressed metric approach \cite{Agullo:2012sh,Agullo:2012fc,Agullo:2013ai} and the hybrid approach \cite{Fernandez-Mendez:2012poe,Fernandez-Mendez:2013jqa,Gomar:2014faa,Gomar:2015oea,Martinez:2016hmn,ElizagaNavascues:2020uyf}. 
In both of these approaches, the background spacetime is loop quantized using the $\bar \mu $ scheme in LQC \cite{Ashtekar:2006wn} while the linear perturbations are Fock quantized on the loop quantum homogeneous spacetime. While the quantum theory of both the approaches has various interesting elements which have been rigorously studied \cite{Agullo:2012fc,Fernandez-Mendez:2013jqa,Gomar:2014faa,Gomar:2015oea}, at a practical computational level both rely on an effective background spacetime encoding quantum gravity effects. In particular, at a  phenomenological level both of the approaches make use of the effective LQC in the background to compute the primordial power spectrum which turns out to be almost scale-invariant for the observable modes.

The dressed metric and the hybrid approaches follow different procedures to reach the Mukhanov-Sasaki equation capturing 
quantum gravity effects via the effective background spacetime, and seemingly have various differences including the way constraints are implemented. The classical theory of dressed metric approach is based on the results for the linear perturbations using Hamiltonian approach in the Langlois' work \cite{Langlois:1994ec}. In particular, the background spacetime is assumed to be a spatially-flat Friedmann-Lema\^itre-Robertson-Walker (FLRW) universe on which the linear perturbations are expanded in terms of the Fourier modes. While Langlois uses spatially-flat gauge to perform most computations in the intermediate steps, final answer is written in terms of the gauge-invariant Mukhanov-Sasaki variable $Q_{\vec k}$ using which the second order Hamiltonian for perturbations in terms of $Q_{\vec k}$ and its conjugate momentum is obtained. This Hamiltonian written in terms of $Q_{\vec{k}}$ serves as the starting point for the quantization in the dressed metric approach. On the other hand, the hybrid approach at the classical level follows the formalism  by Halliwell and Hawking \cite{Halliwell:1984eu}. The original perturbation theory  was understood for a spatially closed universe, and the hybrid approach extends  this formalism to the spatially-flat universe with compact spatial sections. 

When the backreaction of the perturbations on the background dynamics is ignored, which in practice has always been the case so far, 
both approaches yield very similar predictions for the power spectrum given the same initial states in almost all situations for the ultra-violet and intermediate modes at the level of linear perturbations.\footnote{The only known exception in which the initial states in two approaches can not be taken as the same is in one of the modified LQC where the contracting branch turns out to be a quasi de Sitter phase with a Planck-scale cosmological constant  which results in significant differences in phenomenological predictions from two approaches \cite{Li:2019qzr,Li:2020mfi,Li:2021mop}.} The phenomenological difference between these two approaches becomes manifest only near the Planck regime which is captured via the effective mass functions in the Mukhanov-Sasaki equation \cite{ElizagaNavascues:2017avq}. This leads to 
some pertinent questions. 
Given that at a fundamental level both the approaches follow the same strategy -- Fock quantized perturbations over loop quantized background, can the difference between the two approaches at the level of the quantum corrected Mukhanov-Sasaki equation be understood as a result of some quantization ambiguity? Or does the difference between them, at a practical computational level, really arises from adopting different 
methods? Note that there are several examples in the background dynamics of LQC  which reveal that when quantizing the same classical theory, quantization ambiguities can result in dramatic distinctions in the phenomenology (see for eg. \cite{Singh:2013ava,Li:2018opr,Li:2021fmu}). If so, is it possible that 
one can obtain the effective Mukhanov-Sasaki equation used in hybrid approach from a dressed metric like approach? The goal of this manuscript is to answer these questions. Our investigation shows that the difference in two approaches at the computational level arises from using the Mukhanov-Sasaki variable $Q_{\vec{k}}$ versus its rescaled version $\nu_{\vec{k}} = a Q_{\vec k}$, and the non-commuting nature of implementing polymerization in effective spacetime before computing the Poisson bracket (to find the Raychaudhuri equation) or afterwards. We show that if one ignores details of the quantum theory and backreaction effects of perturbations on the background, which is the case in all works so far at the practical computation level to analyze phenomenological implications for perturbations, the ``quantum corrected'' Mukhanov-Sasaki equation with the same effective mass function in the hybrid approach can be easily obtained following a dressed metric like approach in effective spacetime description.

To relate both of these approaches one has to bridge between different strategies and conventions starting at the classical level.  
First, as mentioned above, in the hybrid approach, the perturbation is considered on the background with  compact spatial sections, while in the dressed metric approach there is no such restriction. One can consider spatial hypersurface to be non-compact which then requires an introduction of a fiducial cell to introduce symplectic structure. Secondly, the background metric in the hybrid approach is rescaled  by a constant $\sigma^2=4\pi G/(3l_0)$,  inherited from convention in \cite{Halliwell:1984eu}, with $l_0$ denoting the length of the three-torus spatial manifold. Therefore, when comparing the results from two approaches, one has to take into account this extra scaling constant of the spacetime metric as well. Thirdly, in the hybrid approach, the linear perturbations are expanded in terms of the Fourier modes in the real basis while the dressed metric approach makes use of the complex basis. Fourthly, as mentioned above, the hybrid approach aims at the second order Hamiltonian of the rescaled Mukhanov-Sasaki variable $\nu_{\vec{k}}$ motivated by considerations of unitary quantization of the linear perturbations \cite{Cortez:2010jw,Cortez:2011if} while the dressed metric approach quantizes the second order Hamiltonian in terms of the Mukhanov-Sasaki variable $Q_{\vec k}$. 
Furthermore, when deriving the second order Hamiltonian for $Q_{\vec k}$ or $\nu_{\vec k}$ from the Hamiltonian of the original phase space variables, some canonical transformations are required.  Two different strategies are used in two approaches in order to maintain the system symplectic under the canonical transformations. In the hybrid approach, these canonical transformations are used to redefine the background quantities  \cite{Fernandez-Mendez:2013jqa,Gomar:2014faa,Gomar:2015oea} while in the dressed metric approach the canonical transformations are treated as time-dependent ones as in the Langlois' paper \cite{Langlois:1994ec}. With all the stated differences in the classical aspects of two approaches, one requires correspondence relations between two sets of variables used in two approaches to show they lead to the same classical theory for both background and perturbations. In particular, for consistency the Mukhanov-Sasaki  equations in terms of either $Q_{\vec k}$ or $\nu_{\vec k}$ in both of the approaches are required to have the equivalent classical mass functions at least on the physical solutions of the classical  background dynamics. This has so far not been shown in any work to the best of our knowledge which will be established here to help identify the relationship between the dressed metric and the hybrid approach.

After comparing two approaches at the classical level, let us now briefly discuss how the quantum geometry effects are included in both approaches. When quantizing the background and the linear perturbations in both approaches, the background Hamiltonian is loop quantized in the $\bar \mu$ scheme in LQC \cite{Ashtekar:2006wn} while the linear perturbations are Fock quantized on the quantum background spacetime. At this stage, an important approximation, namely, the Born-Oppenheimer (BO) ansatz is made to decompose the total quantum state into a direct product of the individual quantum state for the background and the perturbations. With this ansatz, the second order Hamiltonian which does not necessarily vanish in the dressed metric approach generates a Schr\"odinger-like equation for the perturbations while the hybrid approach implements the Dirac quantization in which the physical solutions of the background and the perturbations are obtained by requiring the vanishing of the Hamiltonian up to the second order in perturbations. Since the zeroth order Hamiltonian is also constrained to zero, the second order Hamiltonian is thus constrained to vanish in the hybrid approach.  For the practical computations of the power spectrum, the test-field approximation in which the background quantum state is taken to be the Gaussian coherent state is assumed to validate the use of the effective dynamics in both approaches. The effective dynamics is able to present a faithful description of the quantum dynamics in LQC as proved by rigorous numerical simulations \cite{Diener:2014hba,Diener:2014mia, Diener:2017lde,Singh:2018rwa}. Almost all the phenomenological studies in both approaches rely on the effective dynamics, in particular with a kinetic dominated bounce where any effects from the inflationary potential in the bounce regime are ignored.\footnote{An exception is the case of a matter-Ekpyrotic bounce scenario studied for dressed metric approach \cite{Li:2020pww}.} It is  well-known that the Mukhanov-Sasaki equations, when written in terms of $\nu_{\vec{k}}$,  in two approaches differ by their effective mass functions which have the same classical limit but  quite distinct behavior in the Planck regime \cite{ElizagaNavascues:2017avq}. The key question we answer in this manuscript is why the effective mass differs in both approaches and how one can obtain the same mass function of the hybrid approach in the dressed metric approach. It turns out that this difference arises only because of the choice made in writing the second order Hamiltonian for perturbations in one variable or another and some associated polymerization ambiguities. If one chooses to express the Hamiltonian for second order perturbations at effective level in the dressed metric approach in terms of of $\nu_{\vec{k}}$ and uses the same polymerization as used so far in the hybrid approach, the quantum corrected Mukhanov-Sasaki equation turns out to be identical!

In this paper, we first review the classical linear perturbation theory in the formalism used in the literature for each approach. Although the perturbation theory can be done equivalently in different gauges, calculations are more transparent and easier to handle in the spatially-flat gauge. Therefore,  we first present the original Hamiltonian up to the second order in perturbations in the spatially-flat gauge in both approaches. In this gauge, the perturbation of the scalar field is exactly the Mukhanov-Sasaki variable $Q_k$ which greatly simplifies the calculations. We establish the correspondence relations between two sets of canonical variables used in two approaches (see Table \ref{table}) and   explicitly show that the original Hamiltonian in terms of the perturbation of the scalar field in both approaches are exactly equal to each other. Then in order to remove the cross term in this Hamiltonian as well as derive the second order Hamiltonian for perturbations,  one requires to make a canonical transformation. We then show in a generic way that for the same canonical transformation, different strategies used in the dressed metric and the hybrid approach to treat the canonical transformation can lead to the same background and second order Hamiltonian. Specializing our general proof to the spatially-flat gauge, we show that the second order Hamiltonian for the same variable  $\nu_{\vec k}$ in two approaches turns out to be  exactly the same under the correspondence relations in Table \ref{table}. Therefore, when working with the same variable  $\nu_{\vec k}$ , there is no difference at all in the classical mass functions in two approaches.  Besides, we also show that in the dressed metric approach the classical mass function will be different by a term proportional to the background Hamiltonian constraint when working with the variable $Q_{\vec k}$. Moreover, in addition to the spatially-flat gauge, the second order Hamiltonian of $\nu_{\vec k}$ in the hybrid approach is also derived in  the longitudinal gauge \cite{Gomar:2014faa}  as well as the gauge invariant approach \cite{Gomar:2015oea}. We compute the difference between the classical mass functions resulting from using different gauges and find their difference vanishes on the physical solutions of the classical background dynamics.   As a result, we conclude that two approaches are based on the equivalent classical perturbation theory with the same mass function on the physical solutions of the background dynamics.  

In our analysis, for the effective dynamics which incorporates the quantum geometry effects in the background dynamics, we focus on the polymerization of the classically equivalent mass functions. All the classical mass functions are polymerized in a way which is consistent with the polymerization of the background dynamics as initially proposed in the hybrid approach \cite{Fernandez-Mendez:2013jqa}. We find after polymerization, the effective mass functions can be classified into two categories. The first type corresponds to the polymerization of the classical mass function in the dynamical equation for $\nu_{\vec{k}}$ resulting from the Hamiltonian for perturbations written in the Mukhanov-Sasaki variable $Q_{\vec k}$ as in the classical framework of original dressed metric approach. The corresponding effective mass function  after including polymerization capturing quantum gravity effects  (\ref{4.7}) can be  regarded as the one used in the dressed metric approach.\footnote{Note that this expression does not correspond to the same effective mass function as original proposed in original dressed metric formulation \cite{Agullo:2012fc} which turns out to be discontinuous across the bounce. Rather our result agrees with an improved version given in \cite{Li:2019qzr}.} The second type of the effective mass function comes from the polymerization of the classical mass function when working with the rescaled variable $\nu_{\vec k}$ to write the Hamiltonian for perturbations. Its form is given explicitly in (\ref{4.8}) which is usually regarded as the one used in the hybrid approach. We find the difference between these two effective mass functions originates from the non-commutativity of the polymerization and the evaluation of the Poisson bracket. To be specific, the difference comes from the quantum corrections in the modified  Raychaudhuri equation of the scale factor in the effective dynamics of LQC. In the first type, one directly applies the modified  Raychaudhuri equation in the mass functions and thus these quantum corrections are explicitly included. On the other hand,  in the second type, the classical  Raychaudhuri equation of the scale factor is first expressed in terms of the classical canonical phase space variables which is then polymerized. Therefore, the quantum corrections in the  modified Raychaudhuri equation of the scale factor are not included in the second type. One important lesson from our analysis is that with respect to these two different effective mass functions there is no reason in principle at least at the level of effective dynamics  to prefer one over the other considering both of them are coming from the same way of polymerization of the classically equivalent mass functions. In this sense, the difference between the dressed metric and the hybrid approach amounts to some quantization ambiguities which are prevalent among many bottom-up approaches.

The manuscript is organized as follows. In Sec. \ref{sec:review}, we briefly review the classical formulation of the linear perturbation theory in the dressed metric and the hybrid approach. Since two approaches follow Langlois \cite{Langlois:1994ec} and Halliwell \& Hawking \cite{Halliwell:1984eu} respectively, we focus on the classical aspects from the latter while adapting the conventions used in dressed metric and hybrid papers. Following the notations used in each approach, we  present the second order Hamiltonian for the linear perturbation of the scalar field in the spatially-flat gauge from which deviations between two approaches start to emerge.  In Sec. \ref{sec:comparison}, we explicitly show that although the canonical transformations required to obtain the second order Hamiltonian in the dressed metric and the hybrid approach are treated in different strategies, they can lead to the same Mukhanov-Sasaki equations with the equivalent mass functions on the physical solutions of the background dynamics. Then in Sec. \ref{sec:effective dynamics}, we point out that  with the test-field approximation, these classically equivalent mass functions can be polymerized into two different effective mass functions which are typically used in the dressed metric and the hybrid approach. Thus the difference between two effective mass functions are essentially due to the choice of variable used for the Hamiltonian and associated quantization ambiguities and it is easy to obtain the effective mass function of the hybrid approach using dressed metric like approach. In Sec. \ref{sec:conculsions}, we summarize our main results. In the following, we use the Planck units $\hbar=c=1$ and keep Newton's constant $G$ implicit in the constant $\kappa$ with $\kappa=8 \pi G$.

\section{Classical aspects of linear perturbation theory in the dressed metric and the hybrid approach}
\label{sec:review}
\renewcommand{\theequation}{2.\arabic{equation}}\setcounter{equation}{0}

In this section, we briefly review the classical linear perturbation theory used in the dressed metric and the hybrid approach. The classical formulation in these schemes is based on the work of Langlois \cite{Langlois:1994ec} and Halliwell \& Hawking \cite{Halliwell:1984eu} respectively, which we refer the reader for more details. While our focus in this manuscript is to compare the dressed metric and the hybrid approaches, it is insightful in this section to follow details of the classical theory directly from the Langlois' and Halliwell \& Hawking's works. Our goal will be to reach an important juncture in the calculation in those approaches which serves as a point of departure when making canonical transformation to obtain the final form of the Hamiltonian for the Mukhanov-Sasaki variable. Note that while presenting details below we follow the Langlois' and Halliwell \& Hawking's approach but adapt them as needed for dressed metric and hybrid approaches. An example is the use of fiducial cell which is absent in Langlois' approach. Another example is the use of a spatially-compact spatially-flat 3-manifold in Halliwell \& Hawking's approach unlike the spatially-closed manifold in their original paper.   
These two approaches use different sets of variables and Fourier bases to express linear perturbations in the $k-$space. In addition, Halliwell \& Hawking use an overall rescaling of the spacetime metric which is also the case in the hybrid approach. To establish a transparent relationship between the dressed metric and the hybrid approach, a correspondence relation between two sets of variables used in two approaches is needed. Besides, in order to compare the classical aspects of two approaches in a straightforward way, we present the second order Hamiltonian of the perturbed scalar field in the spatially-flat gauge in both approaches. This second order Hamiltonian turns out to take exactly the same form in two approaches under the correspondence relation given in this section.

\subsection{Classical theory following Langlois' approach}

The dressed metric approach is based on Hamiltonian formulation of the classical perturbation theory initially worked out by Langlois \cite{Langlois:1994ec} in which the perturbation theory for a spatially-flat FLRW universe filled with an inflaton field  was studied. In the original work by Langlois, the lapse and shift are treated as the Lagrangian multipliers so that the classical phase space\footnote{The perturbation theory in the Hamiltonian framework can also be formulated in the extended phase space where the lapse and shift are treated as dynamical variables, see for example \cite{Giesel:2018opa, Giesel:2018tcw}.} in the ADM formalism are only composed of the following degrees of freedom: $\Gamma$=$\{ \gamma_{ij},\pi^{ij},\Phi,\pi_\Phi \}$, where $\gamma_{ij}$ and $\pi^{ij}$ denote the three metric and its conjugate momentum and $(\Phi,\pi_\Phi)$ represent the inflaton field and its conjugate momentum respectively. The indices $i,j$ denote the spatial indices which run from 1 to 3. The Poisson brackets between these canonical variables are the standard ones 
\bqn
\lb{dressed Poisson}
\{\Phi(x),\pi_\Phi(y)\}&=&\delta^3\left(x-y\right),
\{\gamma_{ij}(x),\pi^{kl}(y)\}=\frac{1}{2}\left(\delta^k_i\delta^l_j+\delta^l_i\delta^k_j\right)\delta^3\left(x-y\right).
\eqn
Correspondingly, the total action of the system can be written as 
\bq
\lb{dressed action}
S=\int d^4x \left(\pi_\Phi \dot \Phi+\pi^{ij}\dot \gamma_{ij}-N\mathcal H-N^i \mathcal H_i\right),
\eq
where $N$ and $N^i$ denote the lapse and the shift respectively. Besides, the scalar and vector constraints are given explicitly by \cite{Langlois:1994ec}
\bqn
\lb{scalar}
\mathcal H&=&\frac{2 \kappa}{\sqrt{\gamma}}\left(\pi^{ij}\pi_{ij}-\frac{\pi^2}{2}\right)-\frac{\sqrt \gamma}{2\kappa}R+\frac{\pi^2_\Phi}{2\sqrt \gamma}+\sqrt \gamma U+\frac{\sqrt \gamma}{2}\partial_i\Phi\partial^i\Phi,\\
\lb{vector}
\mathcal H_i&=&-2\partial_k\left(\gamma_{ij}\pi^{jk}\right)+\pi^{jk}\partial_i \gamma_{jk}+\pi_\Phi \partial_i \Phi,
\eqn
where $\kappa=8\pi G$ and $\gamma$ denotes the determinant of the three-metric, $R$ and $U$  stand for  the intrinsic Ricci scalar and the potential of the scalar field. The phase space variables can then be decomposed into the background sector and the perturbation sector as
\bqn
\lb{dressed decomposition}
\Phi&=&\bar\phi(t)+\delta\phi(t,\vec x),\quad \quad 
\pi_\Phi=\bar \pi_\phi(t)+\delta \pi_\phi(t,\vec x),\nb\\
\gamma_{ij}&=&\bar \gamma_{ij}(t)+\delta \gamma_{ij}(t,\vec x), \quad
\pi^{ij}=\bar \pi^{ij}(t)+\delta \pi^{ij}(t,\vec x),
\eqn
where the barred quantities stand for the background variables,  $\delta\phi$, $\delta \pi_\phi$ etc are the perturbations. 
For the spatially-flat FLRW background, the background variables for the geometrical sector can be parameterized as 
\bq
\lb{dressed background metric}
\bar \gamma_{ij}=a^2 \delta_{ij}, \quad \quad \bar \pi^{ij}=\frac{\pi_a}{6a}\delta^{ij},
\eq
where $a$ is the scale factor and $\pi_a$ is its conjugate momentum. As a result,  the homogeneous sector of the phase space consists of four variables which are $\{a, \pi_a, \bar\phi, \bar \pi_\phi\}$. If the spatial manifold is non-compact, one needs to introduce a fiducial cell ${\cal V}$ with volume $ V_o$ with respect to the fiducial metric to define the symplectic structure.  Assuming the perturbations $\Gamma_1=\{\delta\phi(t,x), \delta \pi_\phi(t,x), \delta \gamma_{ij}(t,x), \delta \pi^{ij}(t,x) \}$ are purely inhomogeneous, it is straightforward to find that
\bq
\lb{dressed symp}
\int d^4x \left(\pi_\Phi \dot \Phi+\pi^{ij}\dot \gamma_{ij}\right)=\int dt  V_o \left(\bar \pi_\phi \dot{ \bar\phi}+\pi_a\dot a\right) +\int d^4x\left( \delta \pi^{ij}  \delta \dot{\gamma}_{ij}+\delta \pi_\phi\delta \dot \phi\right),
\eq
which implies $\{a, \pi_a\}=\{\bar \phi,\bar \pi_\phi\}=1/V_o$.

Using (\ref{dressed decomposition}), the scalar and vector constraints in (\ref{scalar})-(\ref{vector}) can also be expanded up to the second order in perturbations as
\bq
\mathcal H=\mathcal H^{(0)}+\mathcal H^{(1)}+ \mathcal H^{(2)},\quad \quad \mathcal H_i= \mathcal H^{(1)}_i+ \mathcal H^{(2)}_i,
\eq
with $\mathcal H^{(0)}$ standing for the zeroth order scalar constraint,  $ \mathcal H^{(1)}$ and $ \mathcal H^{(1)}_i$  stand respectively for the first order scalar and vector constraints, and similarly $ \mathcal H^{(2)}$ and $ \mathcal H^{(2)}_i$ stand for the second order scalar and vector constraints. The zeroth order scalar constraint is given explicitly by 
\bq
\lb{dressed background}
\mathcal H^{(0)}=-\frac{\kappa \pi^2_a}{12 a}+\frac{\bar \pi^2_\phi}{2a^3}+a^3 U,
\eq
while the exact forms of the first/second order scalar constraint and the first order vector constraint in terms of $\delta \gamma_{ij}$ and $\delta \pi^{ij}$ can be found in \cite{Langlois:1994ec}.\footnote{See Eqs. (19), (20) and (50) in \cite{Langlois:1994ec}. Besides the second order vector constraint is not required for the calculation of the power spectrum so we will ignore it in the following analysis.} As a result, the Hamiltonian of the background and the perturbations turns out to 
\bq
\lb{dressed ham}
\bold H=\int d^3x \left(N\mathcal H+N^i \mathcal H_i\right)= N  V_o  \mathcal{H}^{(0)}+N\int d^3x  \mathcal H^{(2)},
\eq
where we have chosen $N^i=0$. The role of the first order scalar and vector constraints is to impose constraints on the perturbations $\delta \gamma_{ij}$ and $\delta \pi^{ij}$.
To decouple the scalar, vector and tensor modes in the perturbations, it is convenient to work in the momentum space.  The perturbations can be expanded in terms of the Fourier modes with respect to a finite fiducial cell \cite{Agullo:2017eyh} or in the limit of infinite spatial sections \cite{Artigas:2021zdk}. Since the linear perturbation in the hybrid approach is analyzed in the compact spatial sections, in order to bring a closer relationship between two approaches, we consider a finite fiducial cell.  Taking the perturbed scalar field $\delta \phi(t,\vec x)$ as an example, it can be expanded in the Fourier series as 
\bq
\lb{Fourier series}
\delta \phi(t,\vec x)=\sum_{\vec k}\delta \phi_{\vec k}(t)e^{i\vec k \cdot \vec x },
\eq
with its Fourier coefficients given by 
\bq
\delta \phi_{\vec k}(t)=\frac{1}{ V_o}\int d^3x\delta \phi(t, \vec x)e^{-i\vec k \cdot \vec x },
\eq
where the non-zero wavevector takes the discrete value $\vec k=\frac{2\pi}{l_0}\vec n$ with  $\vec n=(n_x,n_y,n_z)\in \nb{Z}^3$ being any tuple of integers and $\vec n \neq \vec 0$. Besides, the reality condition $\delta \phi(t,\vec x)=\delta \phi^*(t,\vec x)$ requires $\delta \phi^*_{\vec k}(t)=\delta \phi_{-\vec k}(t)$. In the momentum space, the standard Poisson bracket now becomes 
\bq
\{\delta \phi_{\vec k}(t),\delta \pi_{\phi_{\vec k'}}(t)\}=\frac{1}{ V_o}\delta_{\vec k,-\vec k'}.
\eq
The perturbations of the metric variables and their conjugate momenta can be expanded in the Fourier series in the same manner. Then the second order Hamiltonian in (\ref{dressed ham}) should also be transformed into a summation over the Fourier modes in the momentum space. 
 
For any particular wavevector $\vec k$ in the momentum space, we can introduce six orthonormal bases as in \cite{Langlois:1994ec} (see also  \cite{Agullo:2017eyh}).  These bases can be collectively denoted by $A^m_{ij}$, where $m$ runs from 1 to 6. Their inverse can be denoted by $A_m^{ij}$ which satisfy $A_m^{ij}A^n_{ij}=\delta_m^n$. In terms of these bases, the perturbations can be decomposed into  
\bq
\delta \gamma_{ij}=\gamma_m A^m_{ij}, \quad \quad \delta \pi^{ij}=\pi^m A_m^{ij},
\eq
where $\gamma_1$ and $\gamma_2$ correspond to the scalar modes, $\gamma_3$ and $\gamma_4$ to the vector modes and $\gamma_5$ and $\gamma_6$ to the tensor modes. As analyzed in detail in \cite{Langlois:1994ec}, the first order scalar constraint $  \mathcal H^{(1)}$ and the longitudinal part of the first order vector constraint $\mathcal H^{(1)}_i$ turn out to be the functions of the scalar modes only. They can be used to remove two unphysical scalar modes and leave only one physical. While the two transverse components of the first order vector constraint can be used to remove two vector modes completely. As a result, the remaining physical degrees of freedom amount to one scalar mode and two tensor modes. In the following, we will focus on the scalar mode.  In general, in order to derive the second order Hamiltonian for the physical scalar mode, there are two common strategies. The first one is to construct the gauge invariant variable which commutes with both of the first order scalar and vector constraint, the other is to impose some gauge fixing conditions and then work in the reduced phase space. Since the first strategy also involves the construction of the gauge degrees of freedom which commute with the gauge invariant quantities, it is a more tedious calculation than the gauge fixing strategy which is a more efficient way to extract physical Hamiltonian for the physical degrees of freedom. For this reason, we will choose the spatially-flat gauge in the following as is the case in the intermediate steps in \cite{Langlois:1994ec}.

Choosing the spatially-flat gauge, $\gamma_1=\gamma_2=0$, then one can solve for $\pi^1$ and $\pi^2$ from the first order constraints and then substitute them into the second order scalar constraint $\mathcal H^{(2)}$ in (\ref{dressed ham}) in the momentum space. After implementing all these procedures, one can find the Hamiltonian in terms of the perturbed scalar field and its conjugate momentum 
\bq
\lb{dressed ham sf}
\bold H^{(2)}=N  V_o \sum_{\substack{\vec k^+}} \mathcal H^{(2)}_\mathrm{SF},
\eq
where in order to avoid  the double counting of the degrees of freedom, the above summation is done for the $\vec k^+$ modes with its first non-vanishing component of the wavevector being strictly positive  and  the subscript ``SF" stands for the spatially-flat gauge with
\bq
\lb{second ham dressed}
\mathcal H^{(2)}_\mathrm{SF}=\frac{|\delta \pi_{\phi_{\vec k}}|^2}{a^3}-\frac{3\bar\pi^2_\phi}{\pi_a a^4}\left(\delta \phi_{\vec k}\delta \pi_{\phi_{-\vec k}}+\delta \phi_{-\vec k}\delta \pi_{\phi_{\vec k}}\right)+ \aleph |\delta \phi_{\vec k}|^2,
\eq
and
\bq
\aleph=ak^2+\frac{3\kappa \bar\pi^2_\phi}{2a^3}-6a^2U_{,\bar \phi}\frac{\bar \pi_\phi}{\pi_a}+a^3U_{,\bar \phi \bar\phi},
\eq
where we have suppressed the time dependence in variables for brevity. Note the above second order Hamiltonian is the original one which is obtained from the gauge fixing in the spatially flat gauge and thus contains a cross term. It is the starting point where  two different strategies for dealing with the canonical transformation to  obtain the final form of the Hamiltonian for the Mukhanov-Sasaki variable are implemented in two approaches. We will discuss and compare these two strategies in detail in the next section.

\subsection{Classical theory following Halliwell \& Hawking's approach}

The classical linear perturbation theory in the hybrid approach follows Halliwell \& Hawking's approach \cite{Halliwell:1984eu}, but for a spatially-flat FLRW universe with a $\mathbb{T}^3$ topology in which  the four-dimensional globally hyperbolic spacetime is  ADM decomposed into $\mathcal M=\mathcal R\times \mathbb{T}^3$. The four-metric of the manifold is  parameterized in terms of the lapse $N$, the shift $N^i$ and  the three-metric $h_{ij}$.  The compact spatial hypersurface is coordinated by $\theta^i$ ($i=1,2,3$), each of these angular coordinate ranges between $0$ and $l_0$.  In this way, the spacetime metric takes the ADM form 
\bq
\mathrm{d} s^2=-N^2 \mathrm{d} t^2+h_{ij}\left(\mathrm{d}\theta^i+N^i \mathrm{d}t\right)\left(\mathrm{d}\theta^j+N^j \mathrm{d}t\right).
\eq
One can define a fiducial metric $^oh_{ij}$ on the three-torus and choose it to be the standard Euclidean metric. Then any functions defined on the spatial manifold $\mathbb{T}^3$ can be expanded in terms of the eigenfunctions  $\tilde Q_{\vec n, \pm}(\vec \theta)$ of the Laplace-Beltrami operator compatible with the metric  $^0h_{ij}$ \cite{Gomar:2015oea}
\bq
\lb{hybrid basis}
\tilde Q_{\vec n, +}(\vec \theta)=\sqrt 2 \cos\left(\frac{2\pi}{l_0}\vec n \cdot \vec \theta\right),\quad \quad \tilde Q_{\vec n, -}(\vec \theta)=\sqrt 2 \sin\left(\frac{2\pi}{l_0}\vec n \cdot \vec \theta\right),
\eq
 where $\vec n=(n_1,n_2,n_3)\in \nb{Z}^3$ is any tuple of integers with its first non-vanishing component being a strictly positive integer and the corresponding eigenvalue $-\omega^2_n=-4\pi^2\vec n\cdot \vec n/l^2_0$. With the basis (\ref{hybrid basis}), the spacetime metric, the scalar field and their conjugate momenta can be decomposed into \cite{Gomar:2015oea}
\bqn
\lb{hybrid decomposition1}
h_{ij}&=&\bar h_{ij}+\sum_{\vec n,\epsilon}\left(2a_{\vec n,\epsilon}\bar h_{ij} \tilde Q_{\vec n,\epsilon}+6b_{\vec n,\epsilon}\sigma^2 e^{2\alpha}\Delta_{ij} \tilde Q_{\vec n,\epsilon}\right),\\
\pi^{ij}&=&\frac{1}{6\sigma^{2}e^{2\alpha}l^{3}_0}\left(\pi_\alpha{}^oh^{ij}+\sum_{\vec n,\epsilon}\pi_{a_{\vec n,\epsilon}}{}^oh_{ij}\tilde Q_{\vec n,\epsilon}+\frac{3}{2}\pi_{b_{\vec n,\epsilon}}\Delta^{ij}\tilde Q_{\vec n,\epsilon}\right),\\
\lb{lapse and shift}
N&=&\sigma\left(N_0+e^{3\alpha}\sum_{\vec n,\epsilon}g_{\vec n,\epsilon}\tilde Q_{\vec n,\epsilon}\right),\quad \quad N_i=\sigma^2e^{2\alpha}\sum_{\vec n,\epsilon}\frac{k_{\vec n,\epsilon}}{w^2_n}\partial_i\tilde Q_{\vec n,\epsilon},\\
\Phi&=&\frac{1}{\sigma l^{3/2}_0}\left(\bar \varphi+\sum_{\vec n,\epsilon}f_{\vec n,\epsilon}\tilde Q_{\vec n,\epsilon}\right),\quad \quad
\lb{hybrid decomposition2}\pi_\Phi=\frac{\sigma}{l^{3/2}_0}\left(\bar \pi_\varphi+\sum_{\vec n,\epsilon}\pi_{f_{\vec n,\epsilon}}\tilde Q_{\vec n,\epsilon}\right),
\eqn
where the background spatial metric is given by $\bar h_{ij}= {}^oh_{ij}\sigma^2e^{2\alpha}$ with $\sigma^2=4\pi G/(3l^3_0)$ and  $\Delta_{ij}=-\hat n_i\hat n_j+\frac{{}^oh_{ij}}{3}$ with $\hat n_i=n_i/\sqrt{\vec n\cdot \vec n}$. Instead of the scale factor $a$, its logarithm $\alpha=\ln a$ is used as the canonical variable for the background geometric degree of freedom and $\pi_\alpha$ denotes its conjugate momentum. Besides, in order to distinguish from the dressed metric approach, we also use $\bar \varphi$  for the homogeneous component of the scalar field and its conjugate momentum is denoted by $\bar \pi_\varphi$.  In the above decomposition, we only consider the scalar modes which are $a_{\vec n,\epsilon}$ and $b_{\vec n,\epsilon}$ from the geometric sector and $f_{\vec n,\epsilon}$ from the matter sector.  Under this decomposition, the homogeneous background is described by 
\bq
\lb{hybrid background metric}
\mathrm{d} s^2=\sigma^2\left(-N^2_0(t) \mathrm{d} t^2+e^{2\alpha(t)}{}^oh_{ij}\mathrm{d}\theta_i\mathrm{d}\theta_j\right),
\eq
with an overall normalization constant $\sigma^2$ which was first introduced in  \cite{Halliwell:1984eu} for a spatially-closed universe for convenience. It can also be used to remove the dependence of the final form of the Hamiltonian on $l_0$. However, due to this overall constant, the background metric in the hybrid approach is no longer the same as the one in the dressed metric approach which is given in (\ref{dressed background metric}). Therefore, when the results in two approaches are compared, one should also take into account an additional constant $\sigma^2$ in a consistent way. Finally, it is worth noting that in addition to this constant, the difference between two series expansions in (\ref{Fourier series}) and (\ref{hybrid decomposition1})-(\ref{hybrid decomposition2}) is that the former uses the complex basis while the latter uses the real basis.  

Plugging (\ref{hybrid decomposition1})-(\ref{hybrid decomposition2}) into the action (\ref{dressed action}),  one obtains
\bq
\lb{action2}
S=\int dt \left(\pi_\alpha \dot \alpha+\bar \pi_\varphi\dot {\bar\varphi}+\sum_{\vec n,\epsilon}\left(\pi_{a_{\vec n,\epsilon}}\dot a_{\vec n,\epsilon}+\pi_{b_{\vec n,\epsilon}}\dot b_{\vec n,\epsilon}+\pi_{f_{\vec n,\epsilon}}\dot f_{\vec n,\epsilon}\right)-\bold H\right),
\eq
where the Hamiltonian of the background and the perturbations takes the form
\bq
\lb{hybrid ham}
\bold H=N_0 \mathcal H^{(0)}+\sum_{\vec n,\epsilon}\left(N_0 \mathcal H^{(2)}+g_{\vec n,\epsilon}\mathcal H^{(1)}_S+k_{\vec n,\epsilon}\mathcal H^{(1)}_V \right) .
\eq
Here $ \mathcal H^{(2)}$ is second order in perturbations and $\mathcal H^{(1)}_S$ and $\mathcal H^{(1)}_V$ arise from the linear perturbation of the scalar and vector constraint respectively.  Their explicit forms can be found in Eqs. (2.7)-(2.9) in \cite{Gomar:2015oea}. Besides, the background Hamiltonian is given by 
\bq
\lb{hybrid background}
\mathcal H^{(0)}=\frac{1}{2e^{3\alpha}}\left(-\pi^2_\alpha+\bar \pi^2_\varphi+2e^{6\alpha}V(\bar \varphi)\right),
\eq
where we have used $V(\bar \varphi)$ for the potential of the scalar field in order to distinguish it from the potential in the dressed metric approach. Compared with the Hamiltonian (\ref{dressed ham}) in the last subsection, the Hamiltonian (\ref{hybrid ham}) contains two additional terms coming from the perturbation of the lapse and the shift multiplied by the linear perturbations of the scalar and the vector constraint. This is because in (\ref{lapse and shift}), the lapse and the shift are treated as dynamical variables in the extended phase space instead of the purely Lagrangian multipliers.  The linear perturbation of the scalar and the vector constraint still play the role of constraining the physical degrees of freedom in the sector of the scalar modes. In particular, one can apply the spatially-flat gauge by imposing 
\bq
\lb{spatially flat in hybrid}
a_{\vec n,\epsilon}=b_{\vec n,\epsilon}=0,
\eq
and then solve for their conjugate momenta from $\mathcal H^{(1)}_S \approx 0, \mathcal H^{(1)}_V \approx 0$.  Plugging the resulting expressions of the momenta into $\mathcal H^{(2)}$ in (\ref{hybrid ham}), one can finally obtain \cite{Fernandez-Mendez:2013jqa} 
\bq
\lb{second ham hybrid}
\mathcal H^{(2)}_\mathrm{SF}=\frac{e^{-\alpha}}{2}\Big\{e^{-2\alpha}\pi^2_{f_{\vec n,\epsilon}}-6e^{-2\alpha}\frac{\bar\pi^2_\varphi}{\pi_\alpha}f_{\vec n,\epsilon}\pi_{f_{\vec n,\epsilon}}+\left(\omega^2_n e^{2\alpha} + e^{4\alpha}V_{,\bar \varphi\bar \varphi} +9e^{-2\alpha}\bar \pi^2_\varphi-6e^{4\alpha}V_{,\bar \varphi} \frac{\bar \pi_\varphi}{\pi_\alpha}\right)f^2_{\vec n,\epsilon}\Big\}.
\eq
The corresponding Hamiltonian for the perturbations in the spatially-flat gauge is then given by
\bq
\lb{hybrid ham sf}
\bold H^{(2)}=N_0\sum_{\vec n,\epsilon} \mathcal H^{(2)}_\mathrm{SF}.
\eq

Let us now write a correspondence between the classical framework of dressed metric approach and the hybrid approach. 
In Table \ref{table}, we list one-to-one correspondence between two sets of canonical variables used in both the approaches. Since we compare the  Hamiltonian in the spatially-flat gauge, for the perturbations, we only list the correspondence relation between the perturbed scalar field. 
Using these relations,  one can directly show that two Hamiltonian in the spatially-flat gauge, namely (\ref{dressed ham sf}) and (\ref{hybrid ham sf}), turn out to be the same. Similarly, one can also show in a straightforward way that the background Hamiltonian in two approaches coincide with each other.  As a result,  the original Hamiltonian in the spatially-flat gauge for the background and the perturbed scalar field in two approaches are equivalent to each other under the mapping listed in Table \ref{table}. One can perform similar analysis for different gauges  and the equivalence between two approaches would not change with a different choice of the gauge.  
Finally, in order to remove the  cross terms in  the Hamiltonian (\ref{dressed ham sf}) and (\ref{hybrid ham sf}), some canonical transformation is required. In the hybrid approach, this canonical transformation is used to redefine the background variables while in the Langlois' paper, the canonical transformation is treated as a time-dependent one. In the next section, we will show these two strategies lead to the same form of the Hamiltonian up to the second order in perturbations. 
\\

\begin{table}
\caption{Correspondence between different variables used in the dressed metric and the hybrid approach. 
In this table, $\sigma^2=4\pi G/(3l^3_0)$.}
\begin{center}
 \begin{tabular}{||c| c| c||} 
 \hline
 Variable & Dressed metric approach &   Hybrid approach\\ [0.5ex] 
 \hline\hline
fiducial volume & $  V_o$  & $l^3_0$ \\ 
\hline
 wavevector & $\vec k$  & $\frac{2\pi}{l_0}\vec n$ \\ 
 \hline
lapse function & $N$  & $\sigma N_0$ \\ 
 \hline
 background scalar degrees of freedom & $(\bar \phi,\bar \pi_\phi)$  & $(\sigma^{-1} l^{-3/2}_0\bar \varphi, \sigma l^{-3/2}_0\bar \pi_\varphi)$ \\
 \hline
background geometric degrees of freedom & $(a,\pi_a)$ & $(\sigma e^{\alpha},\sigma^{-1}e^{-\alpha}l^{-3}_0\pi_\alpha)$\\
 \hline
perturbed scalar field& $\delta \phi_{\vec k}$ & $\frac{1}{\sqrt 2 \sigma l^{3/2}_0 }\left(f_{\vec n,+}-if_{\vec n,-}\right)$\\
 \hline
 conjugate momentum of the perturbed scalar field& $\delta \pi_{\phi_{\vec k}}$ & $\frac{\sigma}{\sqrt 2  l^{3/2}_0}\left(\pi_{f_{\vec n,+}}-i\pi_{f_{\vec n,-}}\right)$\\
 \hline
 scalar potential & $U(\bar\phi)$ &$\sigma^{-4}l^{-3}_0 V(\bar \varphi)$\\
 \hline
\end{tabular}
\end{center}
\lb{table}
\end{table}

\noindent
{\bf Remark:} As we have discussed above, the linear perturbation theory in both approaches starts from a generic action which can be cast into the form
\bq
\lb{generic action}
S=\int dt \Big\{W^a_p\dot W^a_q+X_{p_l}\dot X_{q_l}-N_0\left(\mathcal H^{(0)}(W^a_q,W^q_p)+\mathcal H^{(2)}(W^a_q,W^q_p,X_{q_l},X_{p_l})\right)\Big\},
\eq
where $(W^a_q, W^a_p)$ denote collectively the original background variables, namely $(a,\pi_a,\bar \phi,\bar \pi_\phi)$ in the classical formulation following Langlois' work \cite{Langlois:1994ec} used in the dressed metric approach and $(\alpha,\pi_\alpha,\bar \varphi,\bar \pi_\varphi)$ in the hybrid approach. Meanwhile, $(X_{q_l},X_{p_l})$ stand for the perturbation variables which are $(\delta \phi_{\vec k}, \delta \pi_{\phi_{\vec k}})$ in the dressed metric approach and $(f_{\vec n,\epsilon},\pi_{f_{\vec n,\epsilon}})$ in the hybrid approach when we take the spatially-flat gauge. $\mathcal H^{(0)}$ and $\mathcal H^{(2)}$ are the Hamiltonian for the background dynamics and the perturbations respectively. To be specific, they are (\ref{dressed background}) and (\ref{second ham dressed}) in the dressed metric approach and (\ref{hybrid background}) and (\ref{second ham hybrid}) in the hybrid approach. The generic form of the action in (\ref{generic action}) is the starting point for further analysis in the next section,

\section{The time dependent canonical transformation versus the redefinition of the background variables}
\label{sec:comparison}
\renewcommand{\theequation}{3.\arabic{equation}}\setcounter{equation}{0}

Following the analysis in the last section, we will first show generically that the different treatments of the canonical transformations in the dressed metric and the hybrid approaches  yield the same form of the Hamiltonian up to the second order in perturbations. Then we will consider a specific example of the spatially-flat gauge in both approaches to show that  resulting mass functions in the Mukhanov-Sasaki equation differ only by a multiple of the zeroth order Hamiltonian which vanishes identically on the physical solutions of the background dynamics. In this way, we conclude that two approaches have the same classical Hamiltonian for the background and the perturbations, up to second order in perturbations, to later include the effects of quantum geometry.

To remove the cross terms in the second order  Hamiltonian for perturbations, one needs to use a different set of the perturbation variables  $(V_{q_n},V_{p_n})$. The transformation from $(X_{q_l},X_{p_l})$ to the new canonical pairs $(V_{q_n},V_{p_n})$ is a canonical one which can be generically shown as  
\bq
\lb{canonical transformation}
X_{q_l}=a^n_l V_{q_n}+b^n_l V_{p_n},\quad \quad X_{p_l}=c^n_l V_{q_n}+d^n_l V_{p_n},
\eq
where the coefficients $a^n_l,b^n_l,c^n_l,d^n_l$ are understood to be the functions of the background variables and satisfy the normalization condition
\bq
\lb{condition 1}
b^n_lc^n_{l'}-a^n_ld^n_{l'}=\delta_{ll'}.
\eq 
Correspondingly, one can compute
\bq
\lb{symp1}
X_{p_l}\dot X_{q_l}=\frac{1}{2}X_{p_l}\dot X_{q_l}-\frac{1}{2}\dot X_{p_l} X_{q_l}+\frac{1}{2}\frac{d}{dt}(X_{p_l} X_{q_l}),
\eq
where the last term is a surface term which will be dropped. %so we will drop it  out from the action. 
Since the coefficients in the canonical transformation (\ref{canonical transformation}) depend explicitly on the background quantities, (\ref{symp1}) can be shown as 
\bq
\lb{symp2}
X_{p_l}\dot X_{q_l}=\delta W^a_p\dot W^a_q+W^a_p\frac{d\delta W^a_q}{dt}+V_{p_n}\dot V_{q_n},
\eq
where we have defined
\bq
\lb{delta}
\delta W^a_p=\left(\frac{1}{2}X_{p_l}\frac{\partial X_{q_l} }{\partial W^a_q}-\frac{1}{2}X_{q_l}\frac{\partial X_{p_l} }{\partial W^a_q}\right),\quad
\delta W^a_q=\left(\frac{1}{2}X_{q_l}\frac{\partial X_{p_l} }{\partial W^a_p}-\frac{1}{2}X_{p_l}\frac{\partial X_{q_l} }{\partial W^a_p}\right).
\eq
In addition, the following identities are used in the intermediate steps
\bq
\lb{3.1}
V_{p_n}=\frac{1}{2}X_{p_l}\frac{\partial X_{q_l} }{\partial V_{q_n}}-\frac{1}{2}X_{q_l}\frac{\partial X_{p_l} }{\partial V_{q_n}},\quad 
V_{q_n}=\frac{1}{2}X_{q_l}\frac{\partial X_{p_l} }{\partial V_{p_n}}-\frac{1}{2}X_{p_l}\frac{\partial X_{q_l} }{\partial V_{p_n}}.
\eq
One can see from the right-hand side of (\ref{symp2}) that in addition to the last term which gives the right symplectic structure for the new canonical variables, we also need to deal with the other two terms  which are second order in perturbations in a proper way.  There are two different strategies to treat these two additional terms, one is to absorb them into the Hamiltonian which amounts to treat the transformation (\ref{canonical transformation}) as a time-dependent canonical transformation \cite{Langlois:1994ec}, the other is to absorb these terms into the redefinition of the background variables as in the hybrid approach \cite{Fernandez-Mendez:2013jqa,Gomar:2014faa,Gomar:2015oea,Martinez:2016hmn}.  

\subsection{Dressed metric approach and the time-dependent canonical transformation}
\lb{sec:dressed metric}
Note that the first two terms on the right-hand side of  (\ref{symp2}) are second order in perturbations, besides the second term can be rewritten into the form 
\bq
\lb{symp3}
W^a_p\frac{d\delta W^a_q}{dt}=-\delta W^a_q \dot W^a_p+\frac{d}{dt}\left(\delta W^a_q W^a_p\right).
\eq
Therefore, one can make use of the Hamilton's equations for the background 
\bq
\lb{background ham}
\dot W^a_q=N_0\Big\{W^a_q,\mathcal H^{(0)}\Big\}=N_0\frac{\partial \mathcal H^{(0)}}{\partial W^a_p},\quad \dot W^a_p=N_0\Big\{W^a_p,\mathcal H^{(0)}\Big\}=-N_0\frac{\partial \mathcal H^{(0)}}{\partial W^a_q}.
\eq
Using (\ref{symp3})-(\ref{background ham}), the first two terms on the right-hand side of  (\ref{symp2})  turn out to be  proportional to the lapse function and thus can be combined with the original Hamiltonian. In this way, the action in terms of the old background variables and the new perturbation variables take the final form
\bq
\lb{3.2}
S=\int dt \Big\{W^a_p\dot W^a_q+V_{p_n}\dot V_{q_n}-N_0\left(\mathcal H^{(0)}+\tilde {\mathcal H}^{(2)}\right)\Big\},
\eq
where the new second order Hamiltonian is given by 
\bq
\lb{new ham1}
\tilde {\mathcal H}^{(2)}=\mathcal H^{(2)}-\delta W^a_p\frac{\partial \mathcal H^{(0)}}{\partial W^a_p}-\delta W^a_q\frac{\partial \mathcal H^{(0)}}{\partial W^a_q},
\eq
where for the last two terms we are supposed to use the definitions in (\ref{delta}) and $(X_{q_l},X_{q_l})$ are treated as functions of the new variables $(V_{q_n},V_{p_n})$. From the action (\ref{3.2}), we learn that the background canonical variables $(W^a_q,W^a_p)$ as well as the new perturbation variables $(V_{q_n},V_{p_n})$ maintain the same symplectic structure as the set of old phase space variables. More specifically, their Poisson brackets satisfy the  standard ones, namely
\bq
\{W^a_q,W^b_p\}=\delta^{ab},\quad\quad \{V_{q_n},V_{p_m}\}=\delta_{mn},
\eq
with all other Poisson brackets vanishing. 
Above ansatz is essentially the same as the one used in the Langlois' paper where the additional terms in the new second order Hamiltonian are explained to come from a time-dependent canonical transformation, namely the Poisson bracket between the generator of the canonical transformation and the background Hamiltonian. The canonical transformation is carried out mainly for the purpose of removing the cross term in the Hamiltonian (\ref{second ham dressed}). Depending on the new phase space variables in use there can be two distinctive cases which can finally yield the same Mukhanov-Sasaki equations with equivalent mass functions on the physical solutions of the classical background dynamics. In the following, we will discuss each case in some detail and compare their mass functions explicitly. 

\subsubsection{The Hamiltonian in  $Q_{\vec k}$ and its  Mukhanov-Sasaki  equation}

The original  dressed metric approach is based on Mukhanov-Sasaki equation obtained from the Hamiltonian for perturbations using  $Q_{\vec k}$.
Our calculation used spatially-flat gauge in which the variable $Q_{\vec k}$ and its conjugate momentum $P_{Q_{\vec k}}$ are related to the old variables via the canonical transformation
\bq
\lb{3.3}
Q_{\vec k}=\delta \phi_{\vec k},\quad \quad P_{Q_{\vec k}}=\delta \pi_{\phi_{\vec k}}-\frac{3\bar \pi^2_\phi}{a\pi_a}\delta\phi_{\vec k}.
\eq
One can then compute the new second order Hamiltonian according to (\ref{new ham1}), which for a given mode $\vec k$ turns out to be 
\bq
\tilde {\mathcal H}^{(2)}=\frac{|P_{Q_{\vec k}}|^2}{a^3}+a\left(k^2+\Omega^2\right)|Q_{\vec k}|^2,
\eq
with 
\bq
\lb{3.4}
\Omega^2=\frac{3\kappa \bar\pi^2_\phi}{a^4}-18\frac{\bar \pi^4_\phi}{\pi^2_a a^6}-12a\frac{\bar \pi_\phi U_{,\bar \phi}}{\pi_a}+a^2U_{,\bar \phi \bar\phi}.
\eq
Note that while one used spatially-flat gauge to reach above expression of $\tilde {\mathcal H}^{(2)}$, it being expressed in terms of gauge-invariant variable. If one would have not assumed spatially-flat gauge in deriving above expression, one would obtain the same result up to terms which vanish on background solution. 

After obtaining the Hamilton's equations for $Q_{\vec k}$ and $P_{Q_{\vec k}}$, one can derive the equation of motion for $Q_{\vec k}$, yielding 
\bq
\ddot Q_{\vec k}+3H\dot Q_{\vec k}+\frac{k^2+\Omega^2}{a^2}Q_{\vec k}=0,
\eq
where $H$ stands for the Hubble rate defined via $H=\dot a/a$ and an overdot denotes the differentiation with respect to the cosmic time $t$. In the dressed metric approach one starts from this Mukhanov-Sasaki equation and includes quantum geometric corrections. 

Switching to the Mukhanov-Sasaki variable $\nu_{\vec k}=aQ_{\vec k}$, we can immediately obtain from above equation 
\bq
\lb{ms1}
\nu^{\prime\prime}_{\vec k}+\left(k^2+\Omega^2-\frac{a^{\prime\prime}}{a}\right)\nu_{\vec k}=0,
\eq
where a prime denotes differentiation with respect to the conformal time $d\eta=dt/a$. From (\ref{ms1}), we can define the time-dependent mass function by 
\bq
\lb{effective mass1}
m^2_{\mathrm{SF}}=\Omega^2-\frac{a^{\prime\prime}}{a}=\frac{3\kappa \bar\pi^2_\phi}{a^4}-18\frac{\bar \pi^4_\phi}{\pi^2_a a^6}-12a\frac{\bar \pi_\phi U_{,\bar \phi}}{\pi_a}+a^2 U_{,\bar \phi \bar\phi}-\frac{a^{\prime\prime}}{a}.
\eq
Note that the mass function is with subscript `SF' denoting spatially-flat gauge signifying that the original derivation leading to this mass function assumed spatially-flat gauge. For any other gauge, the same mass function would hold on the space of physical solutions.

Furthermore, it is straightforward to check that with the help of the classical Hamiltonian constraint (\ref{dressed background}), the effective potential $\Omega^2$ given in (\ref{3.4}) reduces to a function of the potential of the scalar field, namely, 
\bq
\lb{dressed metric potential}
\Omega^2=a^2\left( \mathfrak{f}^2 U\pm2\mathfrak{f} U_{,\bar \phi}+U_{,\bar \phi \bar\phi}\right),
\eq
where $\mathfrak{f}=\sqrt{\frac{24\pi G}{\rho}}\dot {\bar\phi}$ with $\rho=\frac{1}{2}\dot {\bar \phi}^2+U$ denoting the energy density of the scalar field, and the `+'/`-' sign  applies to the expanding/contracting phase. The above expression of $\Omega^2$ was derived in the original dressed metric papers \cite{Agullo:2012fc,Agullo:2013ai}, albeit only with the negative sign,  when the background classical Hamiltonian constraint (\ref{dressed background}) vanishes identically. This form of $\Omega^2$ was used in the numerical analysis in the early literature of the dressed metric approach with $\rho$ determined from effective equations. While the above expression of the effective potential uses classical Hamiltonian constraint, coincidentally it can also be obtained in the same form in the Planck regime using modified Friedmann dynamics in effective background spacetime (see discussion below \eqref{ansatz0}).

On the other hand, an equivalent expression of the classical mass function which is directly valid in the effective dynamics in the Planck regime was proposed in \cite{Li:2019qzr} motivated by the construction in hybrid approach to respect super-selection sectors. Using the classical Friedmann equation in (\ref{dressed metric potential}), one can find immediately that
\bq
\lb{effective potential1}
\Omega^2=a^2\left( \mathfrak{f}^2 U+6H\frac{\dot {\bar \phi}}{\rho}U_{,\bar\phi}+U_{,\bar \phi \bar\phi}\right) .
\eq
This form of the potential $\Omega^2$  has a more transparent form suited for LQC  than (\ref{dressed metric potential}) since it still holds in the Planck regime in the effective LQC when the background quantities in the original expression (\ref{3.4}) are polymerized in a way consistent  with the polymerization of the background dynamics, especially taking into account super-selection sectors in the quantum Hamiltonian constraint. We will come back to this point in detail in Sec. \ref{sec:effective dynamics}. 

\subsubsection{The Hamiltonian in $\nu_{\vec k}$ and its  Mukhanov-Sasaki  equation}

In addition to the variable $Q_{\vec k}$ , starting from (\ref{second ham dressed}), one can directly find the second order Hamiltonian for the variable $\nu_{\vec k}$ and its conjugate momentum  by using the canonical transformation 
\bq
\lb{3.5}
\nu_{\vec k}=a\delta \phi_{\vec k},\quad \quad \pi_{\nu_{\vec k}}=\frac{\delta \pi_{\phi_{\vec k}}}{a}-\frac{3\bar \pi^2_\phi}{\pi_a a^2}\delta \phi_{\vec k}-\frac{a}{6}\kappa \pi_a\delta \phi_{\vec k}.
\eq
Then it is straightforward to find the new second order Hamiltonian under this canonical transformation, which turns out to be  
\bq
\lb{3.6}
\tilde {\mathcal H}^{(2)}=\frac{|\pi_{\nu_{\vec k}}|^2}{a}+\frac{1}{a}\left(k^2+\tilde m^2_{\mathrm{SF}}\right)|\nu_{\vec k}|^2,
\eq
here the corresponding mass function is given explicitly by  
\bq
\lb{effective mass2}
\tilde m^2_{\mathrm{SF}}=-\frac{27\bar \pi^4_\phi}{2\pi^2_aa^6}+\frac{5\kappa \bar\pi^2_\phi}{2a^4}+\frac{9\bar \pi^2_\phi U}{\pi^2_a}-12aU_{,\bar \phi}\frac{\bar \pi_\phi}{\pi_a}-\frac{\kappa^2\pi^2_a}{72a^2}+a^2U_{,\bar \phi\bar \phi}-\frac{\kappa}{2}a^2U.
\eq
Choosing the lapse function $N=a$, we can obtain the Mukhanov-Sasaki equation in terms of $\nu_{\vec k}$, namely, 
\bq
\lb{ms2}
\nu^{\prime\prime}_{\vec k}+\left(k^2+\tilde m^2_{\mathrm{SF}}\right)\nu_{\vec k}=0.
\eq
As compared with the mass function given in (\ref{effective mass1}), one can easily check that the difference turns out to be proportional to  the background Hamiltonian $\mathcal H^{(0)}$ given by (\ref{dressed background}), that is 
\bq
\lb{3.7}
\delta m^2_{\mathrm{SF}}=m^2_{\mathrm{SF}}-\tilde m^2_{\mathrm{SF}}=-\frac{9\bar \pi^2_\phi}{a^3\pi^2_a}\mathcal H^{(0)}\approx 0,
\eq
where we have used Eqs. (\ref{effective mass1}), (\ref{effective mass2}) and the classical equation of motion of $a^{\prime\prime}/a$. This indicates that two mass functions are equivalent on the physical solutions of the background dynamics which requires the vanishing of the zeroth order Hamiltonian constraint. Furthermore, using the Hamiltonian constraint (\ref{dressed background}) and the equation of motion of $\bar \phi$, we can obtain an equivalent expression of the mass function (\ref{effective mass2}) which is frequently used in the literature, namely, 
\bq
\lb{mass function hybrid}
\tilde m^2_{\mathrm{SF}}=-\frac{4\pi G}{3}a^2 \left(\rho-3P\right)+\mathfrak U,
\eq
where $P=\frac{1}{2}\dot {\bar \phi}^2-U$ denotes the pressure of the scalar field and the effective potential $\mathfrak U$ is defined by
\bq
\lb{effective potential2}
\mathfrak{U}=a^2\left(U_{,\bar \phi \bar\phi}+48 \pi G U+6H\frac{\dot {\bar{\phi}}}{\rho}U_{,\bar \phi}-\frac{48 \pi G}{\rho}U^2\right).
\eq
It should be noted that the form of the mass function given in  (\ref{mass function hybrid}) coincides with the one used in the hybrid approach \cite{ElizagaNavascues:2017avq}. Besides, the effective potential (\ref{effective potential2}) turns out to be identical to the one given in (\ref{effective potential1}). Moreover, it can be shown that this form of the mass function remains the same on the constraint surface of the background dynamics in the effective theory of LQC when we properly polymerize the background quantities in the original mass function (\ref{effective mass2}) (see Sec. \ref{sec:effective dynamics}). Finally it is worthwhile to note that in the classical theory, the Raychaudhuri equation in the conformal time takes the form
\bq
\frac{a^{\prime\prime}}{a}=\frac{4\pi G}{3}a^2 \left(\rho-3P\right).
\eq
Therefore, the mass function (\ref{mass function hybrid}) is exactly the same as the one in (\ref{effective mass1}) with $\Omega^2$ given by (\ref{effective potential1}). 

\subsection{The hybrid approach and the redefinition of the background variables}
\lb{sec:hybrid}
In this ansatz, we can absorb the first two terms on the right-hand side of (\ref{symp2}) into the redefinition of the background quantities. To be specific, if one defines the new background quantities
\bqn
\lb{new variables}
\tilde W^a_q&=&W^a_q+\frac{1}{2}X_{q_l}\frac{\partial X_{p_l} }{\partial W^a_p}-\frac{1}{2}X_{p_l}\frac{\partial X_{q_l} }{\partial W^a_p}=W^a_q+\delta W^a_q,\nb\\
\tilde W^a_p&=&W^a_p+\frac{1}{2}X_{p_l}\frac{\partial X_{q_l} }{\partial W^a_q}-\frac{1}{2}X_{q_l}\frac{\partial X_{p_l} }{\partial W^a_q}=W^a_p+\delta W^a_p,
\eqn
then it is straightforward to show that up to the second order in perturbations the redefinition of the background variables in the above also preserves the symplectic structure,  that is \cite{Gomar:2015oea}
\bq
\lb{3.8}
W^a_p\dot W^a_q+X_{p_l}\dot X_{q_l}=\tilde W^a_p\dot {\tilde W}^a_q+V_{p_n}\dot V_{q_n} + \mathcal O(4)
\eq
where $\mathcal O(4)$ denotes  terms which are fourth order in perturbations. 
In order to obtain the second order Hamiltonian in terms of the new background variables and the new  perturbation variables, one should first note that the inverse of (\ref{new variables}) turns out to be 
\bq
W^a_q=\tilde W^a_q-\delta \tilde{W}^a_q+\mathcal O(4),\quad \quad W^a_p=\tilde W^a_p-\delta \tilde{W}^a_p+\mathcal O(4),
\eq
where we have defined 
\bq
\delta \tilde{W}^a_q\equiv \delta W^a_q|_{W^a_q\rightarrow \tilde W^a_q,W^a_p\rightarrow \tilde W^a_p},\quad \quad \delta \tilde{W}^a_p\equiv \delta W^a_p|_{W^a_q\rightarrow \tilde W^a_q,W^a_p\rightarrow \tilde W^a_p},
\eq
so that $\delta \tilde W^a_q$($\delta \tilde W^a_p$) has the same form as $\delta W^a_q$($\delta W^a_p$) and the former is in terms of the new background variables while the latter is of the old background variables. Finally, in terms of the new variables, the background and second order Hamiltonian turn out to be 
\bq
\mathcal H^{(0)}(W^a_q,W^q_p)+\mathcal H^{(2)}(W^a_q,W^q_p,X_{q_l},X_{p_l})=\tilde {\mathcal H}^{(0)}(\tilde W^a_q,\tilde W^q_p)+\tilde {\mathcal H}^{(2)}(\tilde W^a_q,\tilde W^q_p, V_{q_n},V_{p_n})+\mathcal O(4),
\eq
where we have defined
\bqn
\lb{new ham2}
\tilde {\mathcal H}^{(0)}(\tilde W^a_q,\tilde W^q_p)&=&\mathcal H^{(0)}(W^a_q,W^q_p)|_{W^a_q\rightarrow \tilde W^a_q,W^a_p\rightarrow \tilde W^a_p},\nb\\
\tilde {\mathcal H}^{(2)}(\tilde W^a_q,\tilde W^q_p, V_{q_n},V_{p_n})&=&-\delta \tilde{W}^a_q\frac{\partial\tilde {\mathcal H}^{(0)}(\tilde W^a_q,\tilde W^q_p)}{\partial \tilde{W}^a_q}-\delta \tilde{W}^a_p\frac{\partial\tilde{ \mathcal H}^{(0)}(\tilde W^a_q,\tilde W^q_p)}{\partial \tilde{W}^a_p}+\mathcal H^{(2)}|_{W^a_q\rightarrow \tilde W^a_q,W^a_p\rightarrow \tilde W^a_p}.\nb\\
\eqn
As a result, the new background Hamiltonian $\tilde {\mathcal H}^{(0)}(\tilde W^a_q,\tilde W^q_p)$ for the new background variables takes the same form as the old one $\mathcal H^{(0)}(W^a_q,W^q_p)$ for the original background variables while the new second order Hamiltonian acquire two additional terms from the use of the tilded background variables in the original zeroth order Hamiltonian. Moreover, comparing with the Hamiltonian in (\ref{new ham1}), one can immediately find that the new second order Hamiltonian in terms of the tilded phase space variables given in (\ref{new ham2}) has the same form as the one given in (\ref{new ham1}) in terms of untilded background variables. Therefore, up to the second order in perturbations, the forms of the Hamiltonian from two different ansatz coincide with each other. Besides, both ansatz are consistent with symplectic structure. As a result, we can conclude that up to the second order in perturbations, two different strategies in treating the canonical transformation from $(X_{q_l},X_{p_l})$ to $(V_{q_n},V_{p_n})$  lead to the formally same result. 

The redefinition of the background variables is mainly used in the hybrid approach in which the original total Hamiltonian is given in the action (\ref{action2}).  Since the perturbation variables in (\ref{action2}) are not gauge invariant, one then needs to proceed either by choosing a particular gauge or constructing gauge invariant variables to work with. In the literature, both directions have been studied  \cite{Fernandez-Mendez:2013jqa,Gomar:2014faa,Gomar:2015oea}. Therefore, we will only briefly review the underlying procedures and cite the main results below.

\subsubsection{The classical mass function in the  spatially-flat gauge}

As discussed in Sec. \ref{sec:review}, in the spatially-flat gauge (\ref{spatially flat in hybrid}), one can find the second order Hamiltonian in terms of $f_{\vec n,\epsilon}$ and its conjugate momentum in (\ref{second ham hybrid})-(\ref{hybrid ham sf}), where the cross term can be removed by the canonical transformation explicitly given by \cite{Fernandez-Mendez:2013jqa}
\bqn
\tilde f_{\vec n,\epsilon}&=&e^{\alpha} f_{\vec n, \epsilon},\quad \quad \tilde \pi_{f_{\vec n,\epsilon}}=e^{-\alpha}\Big[\pi_{f_{\vec n,\epsilon}}-\left(3\frac{\bar \pi^2_\varphi}{\pi_\alpha}+\pi_\alpha\right)f_{\vec n,\epsilon}\Big],\nb\\
\tilde \alpha&=&\alpha-\frac{1}{2}\left(3\frac{\bar \pi^2_\varphi}{\pi_\alpha}-1\right)\sum_{\vec n, \epsilon}f^2_{\vec n,\epsilon},\quad \quad \tilde \pi_\alpha=\pi_\alpha-\sum_{\vec n, \epsilon}\Big[f_{\vec n,\epsilon}\pi_{f_{\vec n,\epsilon}}-\left(3\frac{\bar \pi^2_\varphi}{\pi_\alpha}+\pi_\alpha\right)f^2_{\vec n,\epsilon}\Big],\nb\\
\tilde \varphi&=&\bar \varphi+3\frac{\bar \pi_\varphi}{\pi_\alpha}\sum_{\vec n,\epsilon}f^2_{\vec n,\epsilon},\quad \quad \tilde \pi_\varphi=\bar \pi_\varphi,
\eqn
the old and new phase space variables can be shown to satisfy (\ref{3.8}) with $(V_{q_n},V_{p_n})=(\tilde f_{\vec n,\epsilon},\tilde \pi_{f_{\vec n,\epsilon}})$. Hence the above transformation is a canonical transformation. Besides, from the correspondence relation in Table \ref{table}, one can find the new variable $\tilde f_{\vec n,\pm}$ is proportional to the real and imaginary part of the  rescaled Mukhanov-Sasaki variable $\nu_{\vec k}$ defined in (\ref{3.5}). It is straightforward to show that in terms of the new variables (the tilded variables) for the background and the perturbations, the second order Hamiltonian takes the form \cite{Fernandez-Mendez:2013jqa}
\bq
\tilde {\mathcal H}^{(2)}=\frac{\tilde \pi^2_{f_{\vec n,\epsilon}}}{2e^{\tilde \alpha }}+\frac{1}{2e^{\tilde \alpha }}\left(\omega^2_n+\tilde m^2_\mathrm{SF}\right)\tilde f^2_{\vec n,\epsilon},
\eq
with the mass function given by
\bq
\lb{effective mass3}
\tilde m^2_\mathrm{SF}=e^{2\tilde \alpha}V_{,\tilde \varphi\tilde \varphi}+\frac{1}{2}e^{-4\tilde \alpha}\left(-\tilde \pi^2_\alpha+30 \tilde \pi^2_\varphi-6e^{6\tilde \alpha}V\right)-\frac{9}{2e^{4\tilde \alpha}}\frac{\tilde \pi^2_\varphi}{\tilde \pi^2_\alpha}\left(3\tilde \pi^2_\varphi-2e^{6\tilde \alpha}V\right)-12e^{2\tilde \alpha}\frac{\tilde \pi_\varphi}{\tilde \pi_\alpha}V_{,\tilde \varphi},
\eq
where the index `SF' as before denotes the spatially-flat gauge.\\

\noindent{}
{\bf Remark:}
The mass function (\ref{effective mass3}) turns out to be exactly the same as the one in (\ref{effective mass2}) using the mapping in Table \ref{table}. Therefore,  redefinition of variables and time dependent canonical transformation  yield the same classical Mukhanov-Sasaki equation with the identical time-dependent mass function when working with the  rescaled variable $\nu_{\vec k}$.

\subsubsection{The classical mass function in the longitudinal gauge}
In addition to the spatially-flat gauge, starting from (\ref{action2})-(\ref{hybrid ham}), one can also choose the longitudinal gauge which can be implemented by choosing the gauge conditions \cite{Fernandez-Mendez:2013jqa,Gomar:2014faa}
\bq
\lb{3.9}
b_{\vec n, \epsilon}=0, \quad \quad \pi_{a_{\vec n,\epsilon}}-\pi_\alpha a_{\vec n,\epsilon}-3\bar \pi_\varphi f_{\vec n, \epsilon}=0.
\eq
Similar to the case of the spatially-flat gauge, one can then proceed by imposing the linear  scalar and vector constraints together with the gauge fixing conditions to remove $a_{\vec n,\epsilon}$, $b_{\vec n,\epsilon}$ and their respective conjugate momentum from the system, leaving only the physical degrees of freedom. The only complication arises due to the fact that in the longitudinal gauge, the tilded variable $\tilde f_{\vec n,\epsilon}$($=e^{\alpha} f_{\vec n, \epsilon}$) is no longer the rescaled Mukhanov-Sasaki gauge invariant variable $\nu_{\vec n ,\epsilon}$ which, regardless of the gauge fixing conditions, is defined by 
\bq
\lb{3.10}
\nu_{\vec n ,\epsilon}=e^{\alpha}\Big[f_{\vec n,\epsilon}+\frac{\bar \pi_\varphi}{\pi_\alpha}\left(a_{\vec n,\epsilon}+b_{\vec n,\epsilon}\right)\Big].
\eq
Since $a_{\vec n,\epsilon}$ does not vanish in the longitudinal gauge one can not simply identify $\nu_{\vec n ,\epsilon}$ with $\tilde f_{\vec n,\epsilon}$. Therefore, (\ref{3.10}) should be tailored to the longitudinal gauge in order to find the second order Hamiltonian in terms of  $\nu_{\vec n ,\epsilon}$ and its conjugate momentum. One can find more details on the canonical transformation and the corresponding redefinitions of the background variables in \cite{Fernandez-Mendez:2013jqa,Gomar:2014faa}. Here we only cite the final result for the  second order Hamiltonian for the Mukhanov-Sasaki variable $\nu_{\vec n ,\epsilon}$ and its momentum, which explicitly takes the form \cite{Gomar:2014faa}
\bq
\tilde {\mathcal H}^{(2)}=\frac{ \pi^2_{\nu_{\vec n,\epsilon}}}{2e^{\tilde \alpha }}+\frac{1}{2e^{\tilde \alpha }}\left(\omega^2_n+\tilde m^2_\mathrm{LG}\right) \nu^2_{\vec n,\epsilon},
\eq
with the corresponding mass function given by
\bq
\lb{effective mass4}
\tilde m^2_\mathrm{LG}=e^{-4\tilde \alpha}\left(19\tilde \pi^2_\varphi-18\frac{\tilde \pi^4_\varphi}{\tilde \pi^2_\alpha}\right)+e^{2\tilde \alpha}\left(V_{,\tilde \varphi\tilde \varphi}-4V-12V_{,\tilde \varphi}\frac{\tilde \pi_\varphi}{\tilde \pi_\alpha}\right)
\eq
with the subscript `LG' denoting the longitudinal gauge. 
One can directly compute the difference between two mass functions (\ref{effective mass3}) and (\ref{effective mass4}) resulting from choosing different gauge fixing conditions, and it  turns out that
\bq\label{gauge-fixing-mass}
\delta m^2=\tilde m^2_\mathrm{SF}-\tilde m^2_\mathrm{LG}=\left(9\frac{\tilde \pi^2_\varphi}{\tilde \pi^2_\alpha}+1\right)e^{-\tilde \alpha }\tilde {\mathcal H}^{(0)}.
\eq
Therefore the difference in the mass functions resulting from different choices of the gauge fixing conditions vanishes on the physical solutions of the background dynamics, implying that the physical predictions are independent of the gauge fixing conditions as generally expected for any classical gauge theory.

\subsubsection{The classical mass function in the gauge invariant approach}

In addition to choosing a particular gauge, there is also a gauge invariant approach to obtain the Mukhanov-Sasaki equation by separating the physical degrees of freedom from the gauge degrees of freedom explicitly in the phase space. In the hybrid approach the detailed analysis was first carried out in \cite{Gomar:2015oea}. The starting point is first to abelianize the linear scalar and vector constraints and then parameterize the space of the inhomogeneous perturbations by the abelianized linear constraints and the Mukhanov-Sasaki variable $\nu_{\vec n ,\epsilon}$. One also needs to find their proper conjugate variables so that the transformation from the old variables to the new ones is canonical. The details of the above procedures can be found explicitly in \cite{Gomar:2015oea}. It turns out that the resulting second order Hamiltonian for the Mukhanov-Sasaki variable takes the form 
\bq
\tilde {\mathcal H}^{(2)}=\frac{ \pi^2_{\nu_{\vec n,\epsilon}}}{2e^{\tilde \alpha }}+\frac{1}{2e^{\tilde \alpha }}\left(\omega^2_n+\tilde m^2_\mathrm{GI}\right) \nu^2_{\vec n,\epsilon},
\eq
with the effective mass given by
\bq
\lb{effective mass5}
\tilde m^2_\mathrm{GI}=e^{-4\tilde \alpha}\tilde \pi^2_\alpha+ e^{2\tilde \alpha}\left(V_{,\tilde \varphi\tilde \varphi}+30V-12\frac{\tilde \pi_\varphi}{\tilde \pi_\alpha}V_{,\tilde \varphi}-72e^{6\tilde \alpha}\frac{V^2}{\tilde \phi^2_\alpha}\right).
\eq
One can compute the difference between the above mass with the one given in the longitudinal gauge, namely (\ref{effective mass4}), leading to 
\bq
\delta m^2=\tilde m^2_\mathrm{LG}-\tilde m^2_\mathrm{GI}=\left(144e^{5\tilde \alpha}\frac{V}{\tilde \pi^2_\alpha}-72e^{2\tilde \alpha}\frac{\tilde {\mathcal H}^{(0)}}{\tilde \pi^2_\alpha}-34e^{-\tilde \alpha}\right)\tilde {\mathcal H}^{(0)}.
\eq
Therefore the mass functions in the spatially-flat gauge, the longitudinal gauge as well as from the gauge invariant approach are equivalent on the physical solutions of the background dynamics. More specifically, they only differ by a term proportional to the background Hamiltonian constraint which vanishes identically as required. Furthermore, the effective mass functions resulting from the redefinition of the background variables are also equivalent to those from a time-dependent canonical transformation on the physical solutions of the background dynamics. As a result, two strategies can lead to the same Mukhanov-Sasaki equation for the scalar perturbations and the classical aspects of the dressed metric and the hybrid approach turn out to be equivalent  as expected up to the second order in perturbations. 

\section{Polymerization and the effective mass function}
\label{sec:effective dynamics}
\renewcommand{\theequation}{4.\arabic{equation}}\setcounter{equation}{0}
As discussed earlier, the quantization of the background and the linear perturbations in the dressed metric and the hybrid approach essentially follows the same broad path. In particular, the homogeneous gravitational sector is loop quantized in the $\bar \mu$ scheme in LQC \cite{Ashtekar:2006wn}, the homogeneous matter sector is quantized in the Schr\"odinger representation while the linear perturbations are Fock quantized. As a result, the kinematic Hilbert space is a tensor product of the individual Hilbert space for each sector, namely  $\mathcal H_\mathrm{kin}=\mathcal H^\mathrm{grav}_\mathrm{kin}\otimes \mathcal H^\mathrm{matt}_\mathrm{kin}\otimes \mathcal F$. In this sense, the quantization in both approaches is carried out in a ``hybrid way" where the homogeneous and the inhomogeneous sectors are quantized by means of different quantization approaches. Besides, for phenomenological studies  both approaches rely on the test-field approximation in which the quantum states used for the homogeneous gravitational  sector are the Gaussian coherent states for which effective spacetime description is an excellent approximation \cite{Diener:2014hba,Diener:2014mia, Diener:2017lde,Singh:2018rwa}.  With the effective dynamics for the homogeneous background and the Fock quantization of the inhomogeneous sector, what really matters for practical purpose of computing effects of quantum geometry on the power spectrum is the polymerization of the mass functions in the Mukhanov-Sasaki equation. As discussed in the last section, these mass functions in the classical Mukhanov-Sasaki equation in the dressed metric and the hybrid approach are equivalent on the physical solutions of the classical background dynamics, and one is thus led to the investigation of the polymerized mass functions in different approaches and different gauges. 

\subsection{The polymerized mass functions in the dressed metric approach}

The classical background Hamiltonian in the dressed metric approach is given in (\ref{dressed background}) which is in terms of the canonical pair $a$ and $\pi_a$, while it is well-known in LQC that a more appropriate set of the variables for loop quantization  in the $\bar \mu$ scheme is $(v,b)$ which are related with $(a,\pi_a)$ via 
\bq
\lb{4.1}
v=a^3V_0,\quad \quad b=-\frac{\kappa \beta \pi_a}{6a^2},
\eq
where $\beta$ is the Barbero-Immirzi parameter whose value is fixed to be $\beta = 0.2375$ for the numerical purpose as in previous works in LQC. 
In order to obtain the effective Hamiltonian for the background dynamics, the following ``thumb rule" for the polymerization of the variable  $b$ is generally used in the classical background Hamiltonian\footnote{Such a ``thumb rule'' is only valid for spatially-flat models quantized in standard loop quantum cosmology. It neither holds for spatially-curved spacetimes \cite{Gupt:2011jh}, nor for other versions of regularized Hamiltonian constraint in LQC \cite{Li:2021mop}.}, namely
\bq
\lb{4.2}
b^2\rightarrow \frac{\sin^2(\lambda b)}{\lambda^2} .
\eq
Here  $\lambda=\sqrt{\Delta}$ with $\Delta = 4 \sqrt{3} \pi \gamma \ell_{\mathrm{Pl}}^2$ denoting the minimal eigenvalue of the area operator in LQG. With this polymerization, the effective Hamiltonian for the background dynamics turns out to be 
\bq
\lb{4.3}
\mathcal H^{(0)}_\mathrm{eff}=-\frac{3 v \sin^2(\lambda b)}{8\pi G \lambda^2 \beta^2}+\frac{ p^2_\phi}{2v^3}+v^3 U,
\eq
where we have chosen $N=1$ and defined $p_\phi=V_0\bar \pi^2_\phi$.
With the help of the effective Hamiltonian, it is  straightforward to derive the modified Friedmann and Raychaudhuri equations in LQC, namely, 
\bqn
\lb{4.4}
H^2&=&\frac{8\pi G}{3}\rho\left(1-\frac{\rho}{\rho_c}\right),\\
\lb{4.5}
\frac{a^{\prime\prime}}{a}&=&\frac{4\pi G}{3}a^2\rho\left(1+2\frac{\rho}{\rho_c}\right)-4\pi G a^2 P\left(1-2\frac{\rho}{\rho_c}\right),
\eqn
where $\rho_c=\frac{3}{8\pi G\lambda^2\beta^2}$ is the maximum energy density in LQC and the prime denotes a derivative with respect to the conformal time. The linear perturbations are thus described as propagating on the  background spacetime whose evolution is governed by the effective equations (\ref{4.4})-(\ref{4.5}). As a result, the mass functions of the Mukhanov-Sasaki equation are supposed to be polymerized as well in order to be consistent with the effective dynamics of the homogeneous background.  As we have discussed in Sec. \ref{sec:dressed metric}, depending on the new set of the variables after performing the canonical transformation, there can be two mass functions in the spatially-flat gauge, i.e. (\ref{effective mass1}) and (\ref{effective mass2}) (the latter equivalent to (\ref{effective mass3}) in the hybrid approach), which correspond to the classical mass functions used in the  original dressed metric and the hybrid approach respectively. Since these mass functions include terms related with negative powers of $\pi_a$, these terms should be polymerized in a way consistent with the polymerization of the background (see also ``Remark" at the end of this section). 

To be specific, in the classical mass functions (\ref{effective mass1}) and (\ref{effective mass2}), one should polymerize both $1/\pi^2_a$  and $1/\pi_a$. Using classical Hamilton's equation for scale factor we find $\pi_a=-6a^2b/\kappa\gamma$. Therefore, $1/\pi^2_a$ can be polymerized in a way consistent with the polymerization of the background dynamics, namely,
\bq
\lb{ansatz0}
\frac{1}{\pi^2_a}=\frac{\kappa^2\gamma^2}{36 a^4 b^2}\rightarrow \frac{\kappa^2\gamma^2\lambda^2}{36 v^{4/3} \sin^2(\lambda b)}=\frac{\kappa}{12v^{4/3}\rho},
\eq
where we have used the background Hamiltonian constraint in the effective spacetime in the last step. Note that the final expression of $1/\pi_a^2$ is coincidentally the same as one would obtain using the classical Hamiltonian constraint. When it comes to the polymerization of $1/\pi_a$, one may simply consider a square root of the above equation, which is then the same as the choice made in  original dressed metric approach using classical constraint \cite{Agullo:2012fc, Agullo:2012sh, Agullo:2013ai}. 
Since the resulting effective potential (\ref{dressed metric potential}) turns out to be discontinuous at the bounce this choice has a serious drawback. Hence, if the initial states of the perturbations are given in the pre-bounce branch, one has to deal with a discontinuity in the  Mukhanov-Sasaki equation at the bounce. This problem was resolved using a continuous extension across the bounce  \cite{Li:2019qzr} which was motivated by the ansatz originally used in the hybrid approach. The idea is to respect the superselection sectors prescribed by the quantum operator of the background Hamiltonian constraint \cite{Fernandez-Mendez:2013jqa} which demands the following polymerization 
\bq
\frac{1}{b}\rightarrow \frac{\lambda \sin(2\lambda b)}{2\sin^2(\lambda b)}=\frac{\lambda \cos(\lambda b)}{\sin(\lambda b)},
\eq
and thus leads to  
\bq
\lb{ansatz2}
\frac{1}{\pi_a}\rightarrow -\frac{H}{2v^{2/3}\rho}.
\eq
In this ansatz, the effective potential takes the form given in (\ref{effective potential1})/(\ref{effective potential2}) which is continuous and well-behaved at all times. Therefore, in the following, we make use of this ansatz when polymerizing the classical mass functions (\ref{effective mass1}) and (\ref{effective mass2}).

With the latter ansatz discussed above and the effective Hamiltonian constraint (\ref{4.3}), the classical mass functions given in (\ref{effective mass1}) and (\ref{effective mass2}) are polymerized into (we drop the subscript `SF' for brevity)
\bqn
\lb{4.7}
m^2_\mathrm{eff}&=&-\frac{4\pi G}{3}a^2\rho\left(1+2\frac{\rho}{\rho_c}\right)+4\pi G a^2 P\left(1-2\frac{\rho}{\rho_c}\right)+\mathfrak{U},\\
\lb{4.8}
\tilde m^2_\mathrm{eff}&=&-\frac{4\pi G}{3}a^2 \left(\rho-3P\right)+\mathfrak{U},
\eqn
where $\mathfrak{U}$ is still given by (\ref{effective potential2}) and we have also used the modified Raychaudhuri equation to obtain $m^2_\mathrm{eff}$. Note that although the form of the effective mass function (\ref{4.8}) remains the same as its classical counterpart (\ref{mass function hybrid}), the background quantities in (\ref{4.7})-(\ref{4.8}) are actually  determined by the modified Friedmann equation instead of the classical background dynamics. Besides, although two mass functions are equivalent in the classical theory, they are no longer so at the level of the effective dynamics. One can compute their difference directly, yielding
\bq
\lb{4.9}
\delta m^2_\mathrm{eff}=m^2_\mathrm{eff}-\tilde m^2_\mathrm{eff}=-\frac{8\pi G}{3}a^2\left(\rho+3P\right)\frac{\rho}{\rho_c},
\eq
which does not vanish on the physical solutions of the effective background dynamics, especially in the Planck regime where the energy density becomes comparable with the maximum energy density in LQC. This difference originates from the modified Raychaudhuri equation in LQC. More specifically, it is a consequence of the fact that the polymerization of the classical equation of motion of $a^{\prime \prime}/a$ is not equal to the equation of motion of $a^{\prime \prime}/a$ from the effective dynamics. In particular, $m^2_{\mathrm{eff}}$ computed from (\ref{effective mass1}) uses the expression of modified Raychaudhuri equation directly. But going  from (\ref{effective mass1}) to (\ref{effective mass2}) requires usage of classical Raychaudhuri equation, which on polymerization results in above expression of ${\tilde m}^2_{\mathrm{eff}}$. The non-commutativity of the polymerization and the Poisson bracket to compute dynamical equations leads to the inequivalent expressions of the effective mass functions. Since  ${\tilde m}^2_{\mathrm{eff}}$ is the mass function used in hybrid approach, its difference from $m^2_{\mathrm{eff}}$ can be seen as an artifact of a quantization ambiguity of at what step to polymerize!

\subsection{The polymerized mass functions in the hybrid approach}

In the classical formulation of the perturbation theory in the hybrid approach, one can derive the Mukhanov-Sasaki equation by using different gauges and even in a gauge invariant approach. All these ansatz can lead to the equivalent mass functions which differ by a term proportional to the classical background Hamiltonian constraint. Besides, we have also shown that in the spatially-flat gauge, the classical mass function in the hybrid approach given in (\ref{effective mass3}) is exactly the same as the one given by (\ref{effective mass2}) in the dressed metric approach. Therefore, at the level of the effective dynamics, when we polymerize these classical mass functions, i.e. (\ref{effective mass3}), (\ref{effective mass4}) and (\ref{effective mass5}), by the ansatz given in (\ref{ansatz0}), the polymerized mass functions remain equivalent on the physical solutions of the effective dynamics of the background spacetime. Moreover, the form of the polymerized mass function is  exactly the same as the one given in (\ref{4.8}) if the additional ansatz in (\ref{ansatz2}) for polymerizing $1/\pi_a$ is employed. Although the equations of motion for  $\nu_{\vec n,\epsilon}$ and $Q_{\vec n ,\epsilon}$ are equivalent in the classical theory, their counterparts in the effective dynamics turn out to be different from one another due to the non-commutativity of the polymerization and the evaluation of the Poisson bracket to obtain modified Raychaudhuri equation.  

In the literature, the difference between the effective mass functions (\ref{4.7}) and (\ref{4.8}) is usually regarded as the major distinction between the dressed metric and the hybrid approach since the corresponding Mukhanov-Sasaki equations with these two effective mass functions are the starting points for the analytical and numerical computations of the primordial power spectra.  It can be shown that they result in predictions of the power spectra which deviate from general relativity near the regime of the characteristic wavenumbers in each approach.  Although two effective mass functions can not be differentiated from each other in the classical regime when the energy density is much less than $\rho_c$, they have rather different behavior in the Planck regime \cite{ElizagaNavascues:2017avq,Wu:2018sbr}. For example, one of the properties which are often mentioned in the literature is that near the bounce which is dominated by the kinetic energy of the scalar field,  $m^2_\mathrm{eff}$ is always negative while  $\tilde m^2_\mathrm{eff}$ is positive. Another typical example which exemplifies the physical consequence of two effective mass functions is in the one of the modified LQC model, which is called mLQC-I in the literature.  Due to the emergent Planck-scale cosmological constant in its contracting phase, different choices of the effective mass functions, namely whether we works with (\ref{4.7}) or  (\ref{4.8}), can greatly affect the choice of the initial states and thus the behavior of the power spectrum in the infrared and intermediate regimes \cite{Li:2019qzr,Li:2020mfi}. Previously, this served as an example to distinguish two approaches which are characterized by two distinct effective mass functions.  However, in the current paper, we tend to regard two distinctive effective mass functions in the dressed metric and the hybrid approach as coming from the polymerization of two forms of the classical mass functions which are equivalent in the classical theory. So at least at the level of the effective dynamics, the difference between the dressed metric and the hybrid approach is no more than the difference due to the quantization ambiguities which may exist in any equivalent formulations of the classical theories. 

We conclude this section with a remark on the necessity of consistency of polymerization in hybrid as well as dressed metric approaches. \\

\noindent
{\bf Remark:} A compelling reason to use the same polymerization for variables in the propagation equation for perturbations as in the 
background Hamiltonian constraint is tied to the independence of mass function (hence phenomenology) on gauge fixing conditions. Consider for example a comparison between the mass functions in the spatially-flat and the longitudinal gauge in the hybrid approach.  In the effective spacetime description, one obtains a generalization of Eq. (\ref{gauge-fixing-mass}) with both of its sides polymerized. If the polymerization in the mass functions is not identical to the background Hamiltonian constraint the equation will not be satisfied in the Planck regime. As a result, the difference of effective mass functions will depend on the choice of gauge fixing used for perturbations. This will be unacceptable as phenomenological predictions will not be independent of gauge fixing.

\section{Summary}
\label{sec:conculsions}

Exploration of phenomenological consequences of quantum geometry in primordial power spectrum in LQC is an important avenue to link non-perturbative quantum geometric effects from LQG to observations. Since LQC has been so far not derived from a cosmological sector of LQG, it is pertinent to understand the robustness of predictions from LQC given different approaches to incorporate quantum gravity effects in cosmological perturbations. The two main approaches are the dressed metric and the hybrid approach. These are `bottom-up' approaches in the sense that they aim to capture quantum gravity effects in perturbations using LQC as the fundamental theory which is used to quantize the background spacetime and Fock quantizing cosmological perturbations. For practical computations, both of them make use of the test-field approximation in which the effective dynamics in LQC is valid throughout the whole evolution of the universe from the Planck to the classical regime and back-reaction effects are neglected. Extensive work has been done in both the approaches in the last few years \cite{Agullo:2015aba,Ashtekar:2016wpi,Agullo:2017eyh,Zhu:2017jew,Zhu:2017onp,CastelloGomar:2017kbo,ElizagaNavascues:2017avq,Wu:2018sbr,Li:2020mfi,Agullo:2021oqk,Ashtekar:2021izi}, and it has been found that unless one is interested in the modes exiting the horizon in the Planck regime, starting from same initial states for perturbation and initial conditions for the background there are negligible differences in predictions for the ultra-violet modes in CMB. While similarities of predictions have been noted in literature, the question of precise sense in which these approaches are related to each other has not been explored. Since at a practical computational level of the primordial power spectrum both the approaches do not utilize underlying quantum theory from LQC and ignore back-reaction effects, in this manuscript we have approached this question at the classical and effective spacetime levels. Since both approaches follow different methodology, including different Fourier basis and conventions, our first task was to create a mapping between the variables used in both approaches. Hence, we first compared the formulations of the classical perturbation theory in two approaches and then focused on the effective dynamics to incorporate the quantum geometry effects in each approach. 

To compare the dressed metric and the hybrid approach, for simplicity we worked with the spatially-flat gauge as the majority of calculation in Langlois' work \cite{Langlois:1994ec} on which the dressed metric approach is based. We found that in the spatially-flat gauge when the scalar modes from the metric perturbations are gauged to vanish, the second order Hamiltonian for the perturbation of the scalar field and its momentum turns out to be identical in two approaches at the classical level. The difference in the classical formulation of the perturbation theory in two approaches comes at the next step when one uses a canonical transformation to removes the cross term in the original Hamiltonian for the perturbed scalar field. In the Langlois' work (and hence in dressed metric approach) the canonical transformation is treated as a time-dependent one without affecting the background variables while in the hybrid approach the canonical transformation is used to redefine the background variables. We have showed explicitly that both strategies maintain the symplectic structure of the system and they also end up with the same form of the background Hamiltonian. More importantly we found that both the approaches lead to the same form of the second order Hamiltonian when the correspondence relations between two sets of variables in two approaches given in Table \ref{table} are taken into account. As a result, both approaches yield the same Mukhanov-Sasaki equation  with the equivalent mass functions and the formulation of the classical perturbation theory in the original background variables in the dressed metric approach turns out to be the same as the one formulated in term of the redefined background variables in the hybrid approach up to the second order in perturbations. It is also expected that the equivalence between two approaches at the classical level does not depend on the chosen gauge, that is, in each approach and any gauges one can finally obtain the same mass function for the Mukhanov-Sasaki equation on the physical solutions of the background dynamics. We verified this in the hybrid approach where classical mass functions resulting from different gauges as well as gauge-invariant approach are equivalent on the background dynamics. 

Apart from different strategies to deal with the  canonical transformation, the dressed metric and hybrid approaches employ different gauge-invariant variables. The dressed metric approach uses Mukhanov-Sasaki variable $Q_{\vec k}$, while the hybrid approach uses its rescaled counterpart $\nu_{\vec k}$.  Although the canonical transformations which can lead to the Hamiltonian for $Q_k$ and $\nu_k$ are different, the resulting classical mass function showing up in the  Mukhanov-Sasaki equation turn out to be equivalent on the physical solution of the classical background dynamics. Therefore, the choice of the perturbation variables in the classical theory would have no effects on the physical predictions since they can only result in the mass functions which differ by a term proportional to the background Hamiltonian constraint. While at the classical level this change in choice of variables is trivial, it turns out that  the main cause of difference between dressed metric and hybrid approaches is precisely this choice at the computational level when quantum gravity effects are included in the background spacetime. 

It is important to note that the equivalent classical mass functions turn out to be non-equivalent when the background and the linear perturbations are quantized. In both of the approaches, while one focuses on the effective dynamics under the test-field approximation in LQC, in addition to the polymerization of the background  Hamiltonian, the classical mass function in the Mukhanov-Sasaki equation also needs to be polymerized in a way consistent with the polymerization of the background spacetime. We show that at the practical computational level, it is only the polymerization of the equivalent classical mass functions that results in the main distinction between the effective mass functions given the choice of $Q_{\vec k}$ versus $\nu_{\vec k}$ used in the dressed metric and the hybrid approach respectively. This difference originates from the non-commutativity between the polymerization and the evaluation of the Poisson bracket. In particular, it is due to the difference of when to evaluate $a^{\prime\prime}/a$ and perform the polymerization. We find in general the polymerization of the  equivalent classical mass functions can lead to two types of the effective mass functions. The first type which is used in the original dressed metric approach comes from working with $Q_{\vec k}$. In this case, the term $a^{\prime\prime}/a$ directly appears in the mass function and hence in the effective LQC one is supposed to use the modified Raychaudhuri equation for  $a^{\prime\prime}/a$. On the other hand, the second type of the effective mass function which is used in the hybrid approach comes from working with $\nu_{\vec k}$ where the classical mass function which is expressed in terms of the phase space variables is polymerized. The extra terms between two types of the effective mass functions are exactly the quantum correction terms coming from the quantum geometry effects in the modified Raychaudhuri equation  of the scale factor in the conformal time. These terms have little impact in the classical regime but  become important in the Planck regime  and change the qualitative behavior of the effective mass function in the Planck regime.

In conclusion, despite their different methodologies and underlying assumptions, the phenomenological differences in predictions between dressed metric and the hybrid  approach in the Planck regime essentially result from the differences in effective mass functions which arises from the polymerization of the different versions of the classical mass functions  which are equivalent at the level of the classical background dynamics.  In fact, if one were to consider the dressed metric approach with using variable $\nu_{\vec k}$ and choosing same polymerization for $1/\pi_a$, at a phenomenological level one obtains an identical description as the hybrid approach. The difference between the two approaches, at a practical computation level, is no more than the choice of variable used to write the Hamiltonian for perturbations and associated polymerization ambiguities!

\section*{Acknowledgments}

We thank Anzhong Wang for comments on the manuscript. 
This work is supported by the NSF grant PHY-2110207, and the National Natural Science Foundation of China (NNSFC) with grant 12005186.

\end{document}